\documentclass[twocolumn,           % Format : preprint, twocolumn
               showpacs,            % Pacs : showpacs, noshowpacs
               nopreprintnumbers,     % Preprint: preprintnumbers,
               aps,                 % Society: ...
               prd,          	    % Journal Style : pra, prb, prc, prd, pre,
               letterpaper,             % Size : a4paper, ...
              groupeaddress,      % Affiliation (Title) : groupedaddress,
               nofootinbib,         % Footnote: footinbib, nofootinbib
               tightenlines,        % Remove additional spaces in a line
               floats,floatfix,      % Floating pictures and tables
               showkeys
               ]{revtex4-1}
               
\usepackage[toc,page]{appendix}
\usepackage{graphicx}% Include figure files
\usepackage{dcolumn}% Align table columns on decimal point
\usepackage{bm}% bold math
\usepackage{amsmath}
\usepackage{amsfonts,amssymb}
\usepackage{soul}
\usepackage{color}
\definecolor{v}{rgb}{0.6, 0.2, 0.8} %comentarios VM
\definecolor{MAGA}{rgb}{0.1, 0.43, 0.75}
\definecolor{jm}{rgb}{0.13, 0.48, 0.64}

\usepackage{xcolor}
\usepackage{enumerate}
\usepackage{float}
\usepackage{subfigure}
\usepackage{ulem}
\makeatletter
\newcommand{\vast}{\bBigg@{3}}
\newcommand{\Vast}{\bBigg@{4}}
\makeatother

\begin{document}

\title{$\Lambda$CDM-Rastall cosmology revisited: constraints from a recent Quasars datasample}

\author{J.A. Astorga-Moreno$^{1}$}
\email{jesus.astorga@cinvestav.mx}
\author{Kyra Jacobo$^2$}
\email{kyrajacobo@gmail.com}
\author{Salvador Arteaga$^2$}
\email{salvador.arteaga@ibero.mx}
\author{Miguel A. Garc\'ia-Aspeitia$^2$}
\email{angel.garcia@ibero.mx}
\author{A.  Hern\'andez-Almada$^3$}
\email{ahalmada@uaq.mx}

\affiliation{$^1$Departamento de F\'isica, Centro de Investigaci\'on y Estudios Avanzados del IPN, Apartado Postal 14-740, 07000, CDMX, M\'exico.}
\address{$^2$Depto. de F\'isica y Matem\'aticas, Universidad Iberoamericana Ciudad de  M\'exico, Prolongaci\'on Paseo de la Reforma 880, M\'exico D. F. 01219,  M\'exico.}
\affiliation{$^3$Facultad de Ingenier\'ia, Universidad Aut\'onoma de 
Quer\'etaro, Centro Universitario Cerro de las Campanas, 76010, Santiago de
Quer\'etaro, M\'exico.}

%-------------------------------------------------------------------------------------------------
%-------------------------------------------------------------------------------------------------
\begin{abstract}
In this paper we study the impact of a recent quasar datasample in the constraint of the free parameters of an extension of general relativity. As a ruler to test, we use Rastall gravity in the context of background cosmology being a simple extension to general relativity. We compare the results from quasars dataset with other known samples such as cosmic chronometers, supernovae of the Ia type, baryon acoustic oscillations, HII galaxies, and also a joint analysis. Results are consistent with the standard cosmological model emphasizing that Rastall gravity is equivalent to General Relativity. According to the constraints provided from the joint sample, the age of the Universe is $\tau_U = 12.601^{+0.067}_{-0.066}$ Gyrs and the transition to an accelerated phase occurs at $z_T=0.620\pm0.025$ in the redshift scale, being only the phase transition consistent with the standard paradigm and having a younger Universe. With the quasars sample, the universe age differs with that expected in $\Lambda$CDM having a result of $\tau_U = 11.958^{+0.139}_{-0.109}$ Gyrs with a transition at $z_T=0.652\pm0.032$ this last consistent with standard cosmology. A remarkable result is that quasars constraints has the capability to differentiate among general relativity and Rastall gravity due to the result for the parameter $\lambda=-2.231^{+0.785}_{-0.546}$. Moreover, the  parameter $j$ under quasars constraints suggests that the cause of the late universe's acceleration is a dark energy fluid different from a cosmological constant.
\end{abstract}
\keywords{Quasars, Rastall gravity, Cosmology.}
%\draft
\pacs{}
%\date{\today}
\maketitle

%%%%%%%%%%%%%%%%%%%%%%%%%%%%%%%%%%%%
\section{Introduction}
%%%%%%%%%%%%%%%%%%%%%%%%%%%%%%%%%%%%
The understanding of the current Universe's acceleration reported in \cite{Perlmutter:1999,Riess:1998} (and confirmed by other observations \cite{Planck:2018}), is crucial in order to have a general framework of the evolution of the Universe. Based on the General Relativity (GR) knowledge it is possible to establish a scenario that describes the cosmology with an outstanding precision. The whole Universe can be simulated assuming the symmetries of homogenity and isotropy in which the line element that fulfils these characteristics is given by the well-known Friedmann-Lemaitre-Robertson-Walker (FLRW) metric. Thus, in order to have a late Universe acceleration, the GR theory demands a fluid whose Equation of State (EoS) takes the form $w<-1/3$ under the FLRW scenario. If we do not demand this, we take the risk of violating the general covariance of Einstein's field equations, having $\nabla^{\mu}T_{\mu\nu}\neq0$ and thus a source term emerges, producing a late acceleration without the  $w<-1/3$ demand. Of course, exists a plethora of DE models where it is not necessary the negative sign in the EoS of the fluid to have a late Universe acceleration (see for instance \cite{DiValentino:2021izs,Motta:2021hvl}  where some of these DE models are explored).

Both approaches (either the addition of fluid or the rupture of the general covariance) to tackle the problem of the Universe's acceleration are valid, but in the first one it is possible to construct the standard cosmology known as $\Lambda$-Cold Dark Matter ($\Lambda$CDM), where the late acceleration is driven by the cosmological constant $\Lambda$. However, despite its success, $\Lambda$CDM is afflicted with several pathologies such as the fine tuning and coincidence problems \cite{Carroll:2000}. Moreover, we have the problem of the $H_0$ tension, which is a discrepancy on its value that appears when we determine its value using Cosmic Microwave Background Radiation (CMB) through the Planck satellite \cite{Planck:2018}, and compare it to the result of local measurements of $H_0$ given by supernovae of the Ia type of the SH0ES collaboration \cite{Riess:2021jrx}. This problem remains open nowadays and multiple hypotheses have arisen. For example, there is a possibility that the tension could be caused by systematics of distance ladder measurements instead of new physics \cite{Efstathiou:2021ocp} (see also \cite{DiValentino:2021izs,Motta:2021hvl} for a review of the  models).

One of the widely investigated modifications to the
GR are the well-known $f(R)$ theories of gravity \cite{fr,fr2,fr3},
where a function of the Ricci scalar appears in the associated Lagrangian\footnote{ Other important papers studied in literature with modifications to GR and robust constraints can be checked in \cite{Ayuso:2020dcu,Ayuso:2021vtj,Gohar:2020bod,Lazkoz:2019sjl}.}. In 1972, following this direction, Rastall's proposal \cite{rast} relaxes the conservation of the Energy Momentum Tensor (EMT) vanishing in flat space-time but not in an expanding universe on cosmological scales \cite{rast2}, represented
through a relation between the EMT 
 and the simplest curvature invariant in a Riemannian manifold: $\nabla^{\mu}T_{\mu\nu}  
\propto\partial_{\mu}R$. In consequence, the term $g_{\mu\nu}R$ is added to Einstein
equations of GR leading to a
simple mathematical generalization, making viable and significant to extend the standard $\Lambda$CDM model since the observational signatures derived from the non-conservation in the dark sector 
in the non-linear regime on intermediate or small scales, is consistent with the cosmological data \cite{l1,l2,l3,l4}. Roughly speaking, it is not clear whether these two modifications act in the Friedmann equation, and, if so, whether these together could address the low redshift tensions. In recent years, the research community has shown interest in the Rastall scenario, resulting in even more contributions to gravitation and cosmology \cite{a,b,c,d}. Thus, Rastall gravity is a modification to GR parameterized with the free parameter $\lambda$ that in the limit $\lambda\to 0$ recovers GR.  The fact that Rastall Gravity is compatible to GR makes this cosmological model a perfect laboratory to study the efficiency of different data samples that help us constraint cosmological theories; and elucidate if data samples point out signals of a possible theory beyond GR. In this paper we will be specifically studying by the use of Quasi-Stellar Objects datasample as a tester of the theory.

Working with the QSO datasample involves studying data taken from Active Galactic Nuclei (AGN), particularly Quasi-Stellar Objects (QSO) as standard candles. In this case, standardization is key for using QSO as a standard candle for testing the Universe's evolution. For instance, the high stochasticity method is the first attempt to standardize QSO. Additionally, a second attempt is through the non-linear relation between the optical/ultraviolet and the X-ray luminosity in quasars \cite{Lusso:2020fax}. Another method is based on the Radius-Luminosity relations which consists on measuring the responses of the emission lines to the continuous variations, thus resulting in a linear relation between the size of the broad line region and the continuum luminosity \cite{peterson2004}.  At present, the optical data span over eight orders of magnitude of luminosity and cover a wide region $0<z<3.36$. Thus, the Radius-Luminosity relations could be standardizable and useful both for studying evolution in cosmology and as a feasible method to constraint free parameters.  Moreover, in recent studies \cite{Bargiacchi:2023rfd,Rezaei:2020lfy} QSO is used combined with other datasamples to constraint diverse forms of DE through the cosmography approach and investigate a possible tension of the flat $\Lambda$CDM model. In these cases, it is possible to elucidate extra information at intermediate redshifts in regions where other datasamples cannot reach.

The outline of the paper is as follows: in Sec. \ref{1} we describe the mathematical background of Rastall gravity where we construct the $\Lambda$CDM-Rastall model ($\Lambda$CDM-R). In Sec. \ref{sec:constraints} we describe the different data samples that help us to constraint the theory and in particular we show a recent quasar sample. Sec \ref{RE} is dedicated to present our results and finally in Sec. \ref{CO} our conclusions and outlooks are presented.

%%%%%%%%%%%%%%%%%%%%%%%%%%%%%%%%%%%%%%
\section{Mathematical background} \label{1}
%%%%%%%%%%%%%%%%%%%%%%%%%%%%%%%%%%%%%%

We start using the Rastall field equation, which can be written in the form \cite{rast}

\begin{equation}
    G_{\mu\nu}=8\pi G\left(T_{\mu\nu}+\frac{\lambda}{4(1-\lambda)}g_{\mu\nu}T\right),
\end{equation}
where $G_{\mu\nu}$ is the Einstein tensor, $G$ is the Newton gravitational constant, $T_{\mu\nu}$ is the energy-momentum tensor, $\lambda$ is the Rastall parameter, recovering GR when $\lambda\to0$.

If we impose homogeneity and isotropy in our line element given by the Friedmann-Lemaitre-Robertson-Walker conformally flat metric (FLRW) $ds^2=-dt^2+a(t)^2(dr^2+r^2d\Omega^2)$, where $d\Omega^2=d\theta^2+\sin^2\theta d\varphi^2$ and also consider the energy-momentum tensor for a perfect fluid: $T_{\mu\nu}=pg_{\mu\nu}+(\rho+p)u_{\mu}u_{\nu}$ , which contains pressure ($p$), energy density ($\rho$) and cuadri-velocity ($u_{\mu}$), we have the dimensionless Friedmann equation under Rastall approach as

\begin{eqnarray}
E^2(z)=
\Omega_{0r}(z+1)^4+\frac{1}{4}\Omega_{0m}\left[\frac{3\lambda-4}{\lambda-1}\right](z+1)^{3/\beta}+\nonumber\\
\left(1-\Omega_{0r}-\frac{1}{4}\Omega_{0m}\left[\frac{3\lambda-4}{\lambda-1}\right]\right), \label{eq:E(z)}
\end{eqnarray}
where the flatness condition is assumed and $\beta\equiv1-\lambda/4(1-\lambda)$. Just as in the previous Friedmann equation obtained with GR, this one also contains radiation, matter and CC, which means that the Rastall theory adds on the new parameter $\lambda$. Notoriously, radiation is decoupled (it does not depend on the free parameter), with Rastall gravity only affecting the matter field. 

On the other hand, it is possible to construct the deceleration parameter which in Rastall frame reads

\begin{eqnarray}
    q(z)=&&\frac{1}{2E^2(z)}\Big\lbrace4\Omega_{0r}(z+1)^4+\frac{3}{4\beta}\left[\frac{3\lambda-4}{\lambda-1}\right]\times\nonumber\\&&(z+1)^{3/\beta}\Big\rbrace-1.
\end{eqnarray}
Moreover, the cosmographic jerk parameter ($j$) can be calculated through the formula
\begin{equation}
    j=q(2q+1)+(1+z)\frac{dq}{dz},
\end{equation}
 where it is possible to observe that $j$ is a function of the deceleration parameter. 

%%%%%%%%%%%%%%%%%%%%%%%%%%%
\section{Data and methodology} \label{sec:constraints}
%%%%%%%%%%%%%%%%%%%%%%%%%%%

We bound Rastall gravity cosmology ($h$, $\Omega_{0m}$,$\lambda$) by using cosmic chronometers (CC), Hydrogen II galaxies (HIIG), Type Ia Supernovae (SNIa), Baryon Acoustic Oscillations (BAO), Quasi-Stellar Objects (QSO) and joint data through a Bayesian Markov Chain Monte Carlo (MCMC) analysis. We use the well known package \texttt{emcee} \cite{Foreman:2013} under Python language.

We establish a configuration to achieve the convergence of the chains using the autocorrelation function, and generate a set of 3000 chains with 250 steps. Additionally, we use a Gaussian prior over $h$ as $h=0.7403\pm 0.0142$ \cite{Riess:2019cxk}, flat priors over $0<\Omega_{0m}<1$ and $-3<\lambda<3$. Thus, the $\chi^2$-function is given by
\begin{equation}\label{eq:chi2}
    \chi_{\rm Joint}^2 = \chi_{\rm CC}^2 + \chi_{\rm HIIG}^2 + \chi_{\rm SnIa}^2  + \chi_{\rm BAO}^2+\chi_{\rm QSO}^2 \,,
\end{equation}
where each term corresponds to the $\chi^2$ function per sample.

%%%%%%%%%%%%%%%%%%%%%%%%%%%%%%%%%%%%%
\subsection{Cosmic chronometers}
%%%%%%%%%%%%%%%%%%%%%%%%%%%%%%%%%%%%%%

For our purposes, the sample of the Hubble parameter contains 31 measurements using differential age method \cite{Moresco:2016mzx,Magana:2017,Moresco:2015cya}) and being cosmological model independent, offers a great test choice not only for the standard model but even for alternative cosmologies as well. Said sample covers a redshift range $0.07<z<1.965$. The $\chi^2$ function can be built as
\begin{equation} \label{eq:chiOHD}
    \chi^2_{{\rm OHD}}=\sum_{i=1}^{31}\left(\frac{H_{th}(z_i)-H_{obs}(z_i)}{\sigma^i_{obs}}\right)^2,
\end{equation}
here, $H_{th}(z_i)$ is the theoretical Hubble parameter represented in Eq. \eqref{eq:E(z)}, meanwhile $H_{obs}(z_i)\pm \sigma_{obs}^i$ is the observational counterpart with its uncertainty at the redshift $z_i$.

%%%%%%%%%%%%%%%%%%%%%%%%%%%%%%%%%%%%%%
\subsection{Type Ia Supernovae (Pantheon$+$)}
%%%%%%%%%%%%%%%%%%%%%%%%%%%%%%%%%%%%%%

In \cite{Scolnic2018-qf, Brout_2022} we can find the  Pantheon$+$ sample, containing 1701 measurements of the luminosity modulus coming from SNIa, representing the largest sample, covering a region $0.001<z<2.26$. Considering that this sample is extracted from 1550 distinct SNIa, we build the $\chi^2$ function as \cite{Conley_2010}
\begin{equation}\label{eq:chi2SnIa}
    \chi_{\rm SnIa}^{2}=a +\log \left( \frac{e}{2\pi} \right)-\frac{b^{2}}{e},
\end{equation}
where
\begin{eqnarray}
    a &=& \Delta\boldsymbol{\tilde{\mu}}^{T}\cdot\mathbf{Cov_{P}^{-1}}\cdot\Delta\boldsymbol{\tilde{\mu}}, \nonumber\\
    b &=& \Delta\boldsymbol{\tilde{\mu}}^{T}\cdot\mathbf{Cov_{P}^{-1}}\cdot\Delta\mathbf{1}, \\
    e &=& \Delta\mathbf{1}^{T}\cdot\mathbf{Cov_{P}^{-1}}\cdot\Delta\mathbf{1}. \nonumber
\end{eqnarray}
Also, $\Delta\boldsymbol{\tilde{\mu}}$ is the vector of the difference between the theoretical distance modulus and the observed one, $\Delta\mathbf{1}=(1,1,\dots,1)^T$, keeping in mind that the super-index $T$ denotes
the transpose of the vectors
and $\mathbf{Cov_{P}}$ is the covariance matrix formed by adding the systematic and statistic uncertainties. The theoretical counterpart of the distance modulus is estimated by
\begin{equation}
    m_{th}=\mathcal{M}+5\log_{10}\left[\frac{d_L(z)}{10\, {\rm pc}}\right],
\end{equation}
$\mathcal{M}$ represents a nuisance parameter which has been marginalized by Eq. \eqref{eq:chi2SnIa}. Finally, $d_L(z)$ is the luminosity distance, computed through
\begin{equation}\label{eq:dL}
    d_L(z)=(1+z)c\int_0^z\frac{dz^{\prime}}{H(z^{\prime})},
\end{equation}
where $c$ is the speed of light.

%%%%%%%%%%%%%%%%%%%%%%%%%%%%%%%%%%%%%%
\subsection{Baryon Acoustic Oscillations}
%%%%%%%%%%%%%%%%%%%%%%%%%%%%%%%%%%%%%%

The Baryon Acoustic Oscillations signature is the result of the baryons-photons interactions in the recombination epoch and it has proven useful for bounding model parameters. 

We use the latest measurements reported in \cite{Alam_2021} which consist of 14 uncorrelated data points covering a redshift range of $0.15<z<2.33$. The Rastall cosmology is constrained via the $\chi^2$ function
\begin{equation}\label{eq:bao}
 \chi^2_{{\rm BAO}}=\sum_{i=1}^{14}\left(\frac{\mathcal{M}_{th}(z_i)-\mathcal{M}_{obs}(z_i)}{\sigma^i_{obs}}\right)^2,
\end{equation}
where where $\mathcal{M}_{obs}$ represents the observational measurement and $\mathcal{M}_{th}$ represents the theoretical estimate of the quantities $D_V(z)/r_d$, $D_M(z)/r_d$, $D_H(z)/r_d = c/H(z)r_d$ where
\begin{equation}
    D_V(z) = [z\,D_H(z)\,D_M^2(z)]^{1/3}\,.
\end{equation}
is the dilation scale \cite{Wigglez:Eisenstein2005}, $r_d = r_s(z_d)$ is the size of the sound
horizon at the drag epoch redshift $z_d$ given by
\begin{equation}
    r_d = \int_{z_d}^\infty \frac{c_s(z)dz}{H(z)}\,,
\end{equation}
in which $c_s(z)$ is the sound speed and 
\begin{equation}
    D_M(z) = c \int_0^z \frac{dz'}{H(z')}\,, \label{DM}
\end{equation}
is the comoving angular diameter for a flat cosmology at the redshift $z$, and $c$ is the speed of light. For this work we use $z_d = 1089.80 \pm 0.21$ \cite{Planck:2018}. For more details about this sample, see Table 3 of \cite{Alam_2021}.

%%%%%%%%%%%%%%%%%%%%%%%%%%%%%%%%%%%%%%
\subsection{HII Galaxies}
%%%%%%%%%%%%%%%%%%%%%%%%%%%%%%%%%%%%%%

 This sample is proposed in order to implement a first approach to galaxies at high redshifts in the era of James Webb telescope \cite{Padmanabhan:2023esp} and its comparison to QSO datasamples in future studies. We have 181 measurements coming from HIIG, with their luminosity dominated by young massive burst of star formation, is reported by \cite{GonzalezMoran2019, Gonzalez-Moran:2021drc}. This sample covers a region of $0.01<z<2.6$ and is useful to establish bounds over cosmological parameters due the correlation between the measured luminosity $L$ of the galaxies and the inferred velocity dispersion $\sigma$ of their ionized gas \cite{Chavez2012,Chavez2014,Terlevich2015,Chavez2016}. The $\chi^2$-function is built as
\begin{equation}\label{eq:chi2_HIIG}
    \chi^2_{{\rm HIIG}}=\sum_i^{181}\frac{[\mu_{th}(z_i, {\Theta})-\mu_{obs}(z_i)]^2}{\epsilon_i^2},
\end{equation}
where $\epsilon_i$ is the observational uncertainty measured at $z_i$ having $68\%$ of confidence level. Additionally, the observational distance modulus ($\mu_{obs}$) is expressed
\begin{equation}
    \mu_{obs} = 2.5(\alpha + \beta\log \sigma -\log f - 40.08)\,.
\end{equation}
Here, $\alpha$ and $\beta$ are the intercept and slope of the $L$-$\sigma$ relation and $f$ is the measured flux. 
The theoretical estimate is given as
\begin{equation}
    \mu_{th}(z, \Theta) = 5 \log_{10} \left [ \frac{d_L(z, \Theta)}{1\,{\rm Mpc}}\right] + 25,
\end{equation}
where $d_L$ is the luminosity distance measured in Mpc (see Eq. \eqref{eq:dL}).

%%%%%%%%%%%%%%%%%%%%%%%%%%%%%%%%%%%%%
\subsection{Quasars}
%%%%%%%%%%%%%%%%%%%%%%%%%%%%%%%%%%%%%%

A compilation of intermediate-luminosity quasars is reported by C. Shuo et. al \cite{ShuoQSO:2017}. A total of 120 QSO luminosity measurements covering a redshift region $0.462<z<2.73$. To constrain cosmological model parameters, the $\chi^2$ function is built as
\begin{equation}
    \chi^2_{\rm QSO} = \sum_i^{120}\frac{[\theta_{th}(z_i)-\theta_{obs}^i]^2}{\sigma^2_i},
\end{equation}
where $\theta_i \pm \sigma_i$ is the observational measurement of the angular size and its uncertainty at the redshift $z_i$ and is written as \cite{Sandage:1988}
\begin{equation}
    \theta(z) = \frac{l_m}{D_A(z)}\,,
\end{equation}
and its theoretical counterpart is denoted by $\theta_{th}$. In the latter expression, $D_A$ is the angular diameter distance  $D_A(z) = c/(1+z)\int_0^z dz'/H(z')$ and $l_m$ is an intrinsic length. We use the value $l_m=11.03 \pm 0.25$ pc found in \cite{ShuoQSO:2017}.
%%%%%%%%%%%%%%%%%%%%%%%%%%%%%%%%%%%%%%
\section{Results} \label{RE}
%%%%%%%%%%%%%%%%%%%%%%%%%%%%%%%%%%%%%%

\begin{figure*}
    \centering
    \includegraphics[width=0.6\textwidth]{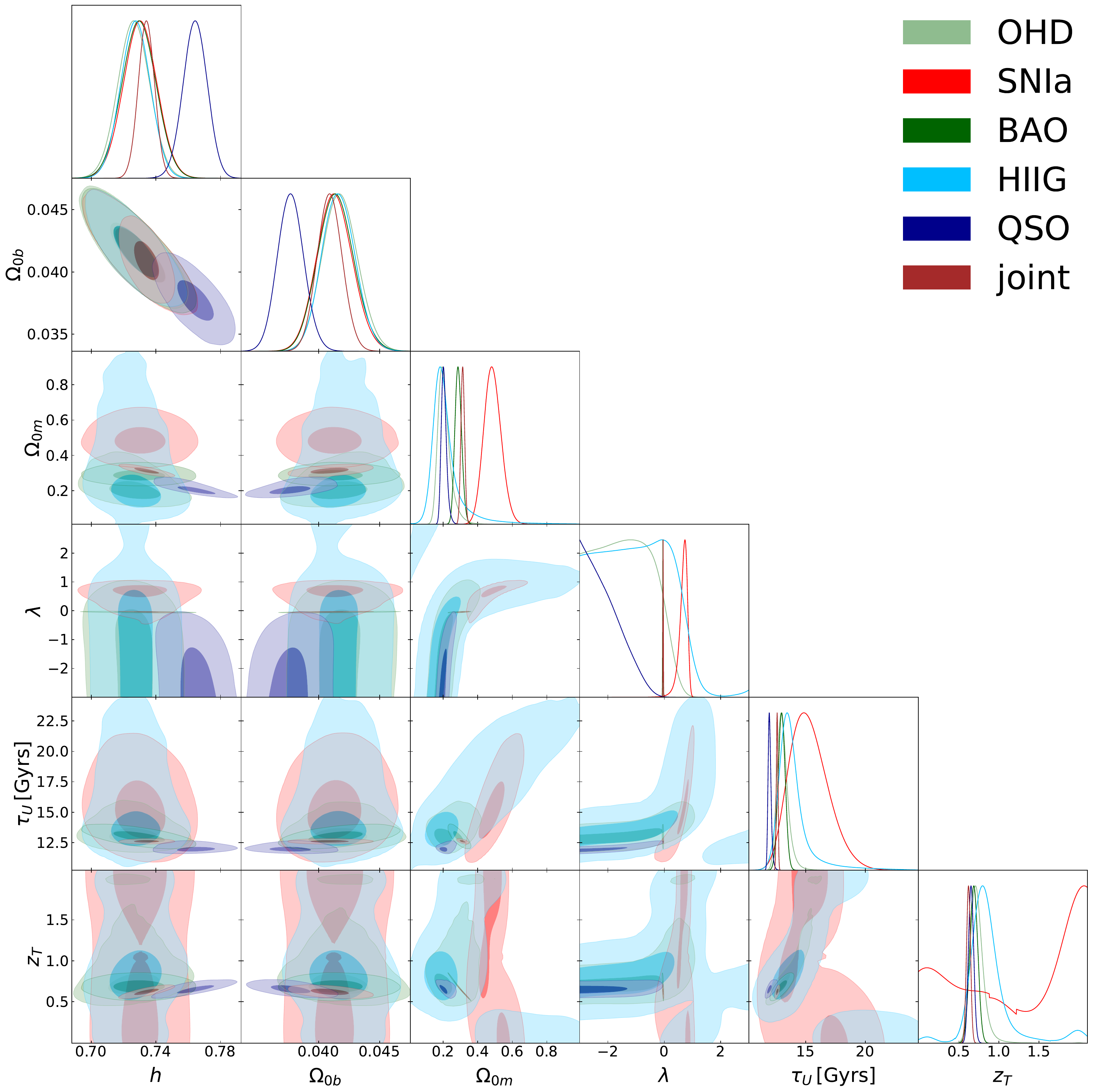}
    \caption{2D contours at $1\sigma$ (inner region) and $3\sigma$ (outermost region) CL for Rastall cosmology.}
    \label{fig:contours}
\end{figure*}

\begin{figure*}
    \centering
    \includegraphics[width=0.32\textwidth]{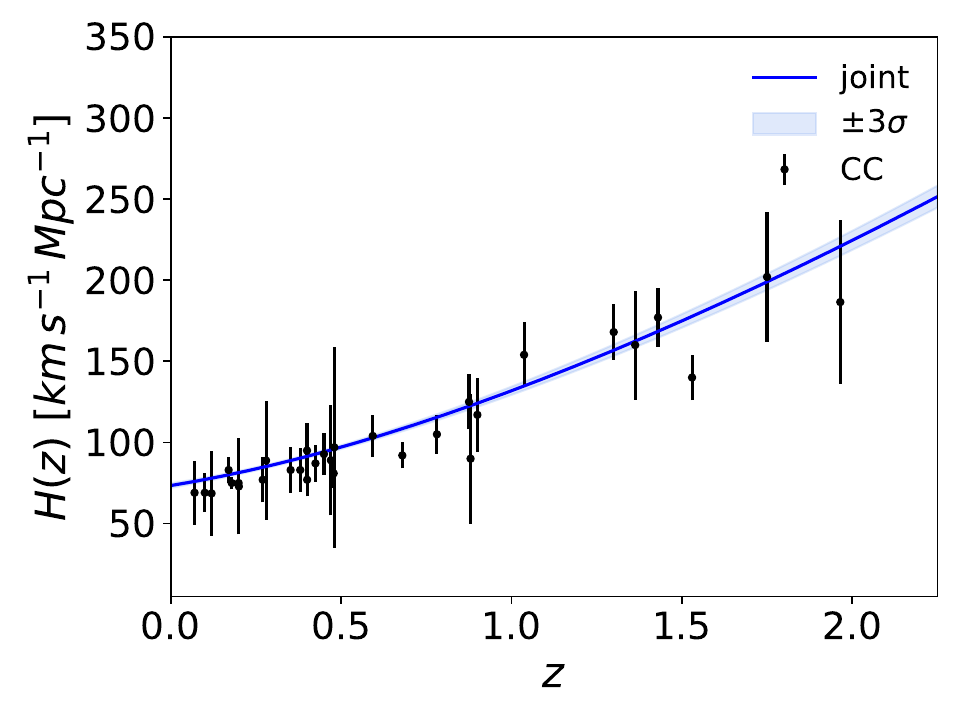}
    \includegraphics[width=0.32\textwidth]{plot_rastall_qz_joint.pdf}
    \includegraphics[width=0.32\textwidth]{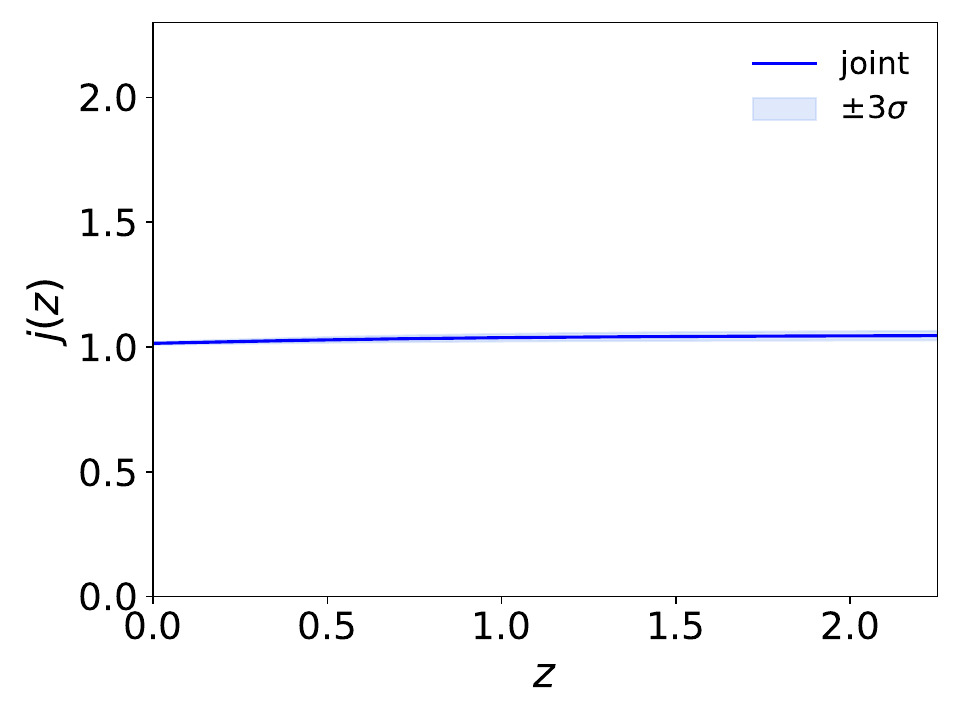} \\
    \includegraphics[width=0.32\textwidth]{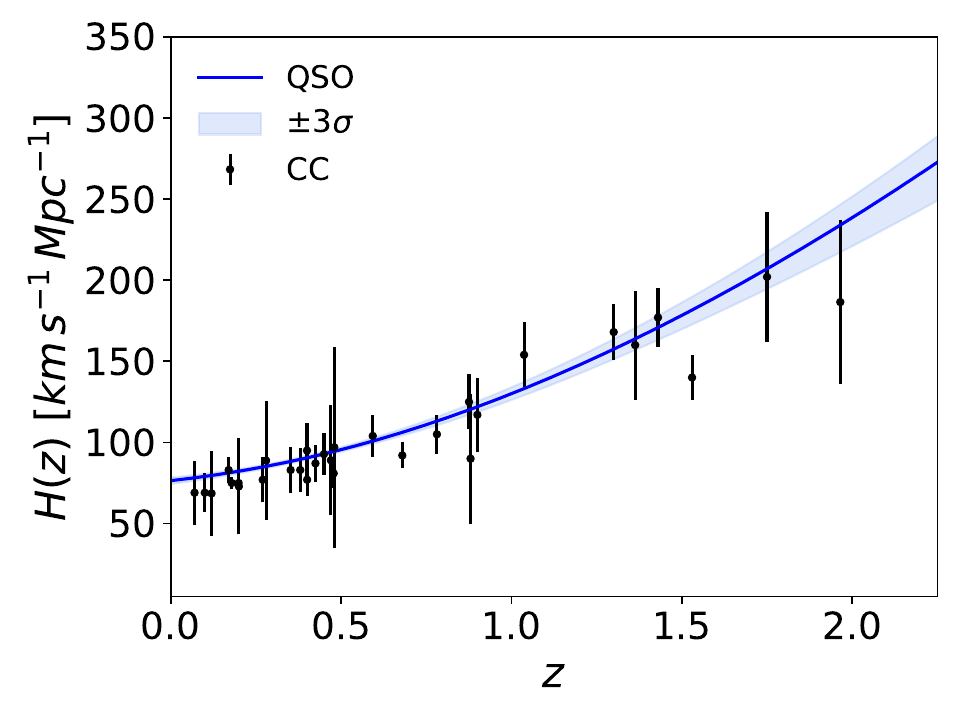}
    \includegraphics[width=0.32\textwidth]{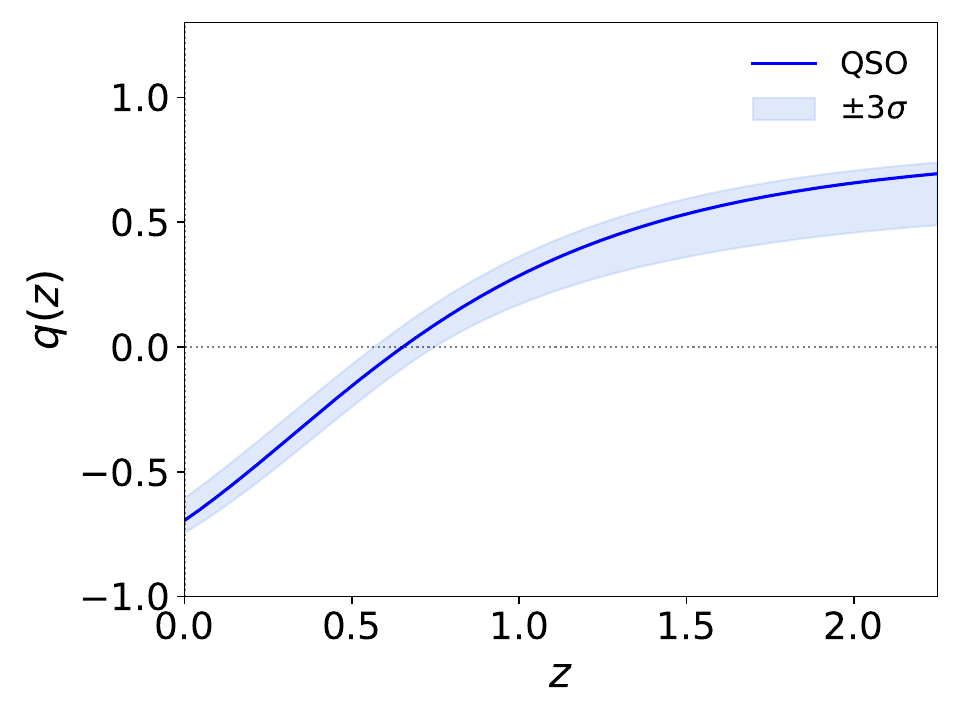}
    \includegraphics[width=0.32\textwidth]{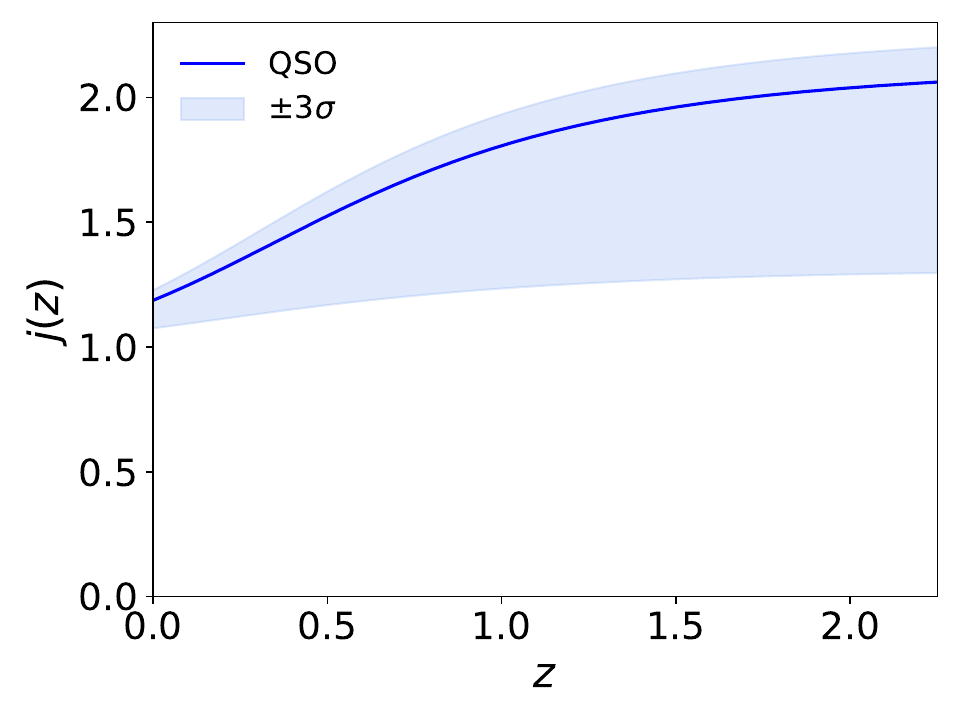}

    \caption{Reconstruction of $H(z)$ (left panel), $q(z)$ (middle panel) and $j(z)$ (right panel) for $\Lambda$CDM-R using joint and QSO sample.}
    \label{fig:cosmography}
\end{figure*}

We use the Rastall gravity in the context of $\Lambda$CDM-R model 
as a tester to explore the efficiency of QSO sample and its interaction with other cosmological samples in a joint analysis. In Fig. \ref{fig:contours} we present the 2D parameter likelihood contours at $68\%$ (1$\sigma$) and $99.7\%$ (3$\sigma$) confidence level (CL), remarking in navy blue the data sample of QSO. Moreover, Table \ref{tab:bf_model} shows the mean values of the parameters and their uncertainties at 1$\sigma$ in  where it is shown QSO and the other cosmological samples (Observational Hubble parameter Data, Supernovae Type Ia, Baryon Acoustic Oscillations, HII Galaxies) together with the joint.
Moreover, Fig. \ref{fig:cosmography} presents the plots of $q(z)$ and $j(z)$ with $3\sigma$ uncertainty bands for joint (top panel) and QSO (bottom panel) datasets,  showing the agreement of $q(z)$ with the standard cosmological model despite the presence of the Rastall parameter as expected. However, for QSO the jerk parameter suggests that the cause of the Universe's acceleration is not a CC (despite the addition of a CC to the Friedmann equations), instead it is a dark energy. We think that this behavior is due to the interaction of $\lambda$ with the CC, mimicking a non constant behavior according to QSO constraints.
In the same vein, we present a $\mathbb{H}0(z)$ diagnostic and its comparison with the standard $\Lambda$CDM model assuming $h=0.6766$ and $\Omega_{0m}=0.3111$ according to \cite{Planck:2018}.

Notice that QSO presents a good performance when it is used to constrain the free parameters of $\Lambda$CDM-R model. In this exercise it is possible to notice that the joint analysis has the capability to elucidate among $\Lambda$CDM and $\Lambda$CDM-R. Indeed, QSO sample remark the most negative value of $\lambda$ suggesting a strong presence of Rastall gravity having $6\sigma$ of difference with $\lambda=0$ which is the standard GR. Notice also that only OHD and HIIG do not distinguish Rastall when it is compared with $\Lambda$CDM having a consistency of $1.3\sigma$. A comparison with Planck satellite results \cite{Planck:2018} can be enunciated as follows: matter density parameter for joint is consistent with Planck, however QSO presents a discrepancy  for the Universe age, Planck predicts an age of $13.78$Gyrs while the joint and QSO samples predict a younger Universe with $12.60$ and $11.95$ Gyrs respectively. The transition to an accelerated phase coincides with the standard paradigm for joint  being $-0.530$ while for the  QSO persists with a slight discrepancy having a value of $-0.695$. Notice that SNIa is the constraint that present stronger discrepancies $-0.273$ in comparison with other datasamples.

\begin{table*}
	\centering
	\caption{Best fit values and their uncertainties at $1\sigma$ of the free parameters of $\Lambda$CDM-R cosmology.}
	\label{tab:bf_model}
	\begin{tabular}{lcccccc} % four columns, alignment for each
    \hline
    Parameter & OHD & SNIa & BAO & HIIG  & QSO & joint\\
    \hline
 $h$  & $0.726^{+0.010}_{-0.010}$  & $0.730^{+0.010}_{-0.010}$  & $0.730^{+0.010}_{-0.010}$  & $0.727^{+0.009}_{-0.009}$  & $0.764^{+0.007}_{-0.007}$  & $0.734^{+0.005}_{-0.005}$  \\ [0.9ex] 
 $\Omega_{0b}$  & $0.042^{+0.001}_{-0.001}$  & $0.041^{+0.001}_{-0.001}$  & $0.041^{+0.001}_{-0.001}$  & $0.042^{+0.001}_{-0.001}$  & $0.038^{+0.001}_{-0.001}$  & $0.041^{+0.001}_{-0.001}$  \\ [0.9ex] 
 $\Omega_{0m}$  & $0.201^{+0.040}_{-0.028}$  & $0.484^{+0.051}_{-0.047}$  & $0.287^{+0.020}_{-0.018}$  & $0.191^{+0.080}_{-0.040}$  & $0.203^{+0.016}_{-0.013}$  & $0.313^{+0.011}_{-0.010}$  \\ [0.9ex] 
 $\lambda$  & $-1.397^{+1.070}_{-1.075}$  & $0.697^{+0.110}_{-0.163}$  & $-0.048^{+0.009}_{-0.009}$  & $-0.999^{+1.315}_{-1.348}$  & $-2.231^{+0.785}_{-0.546}$  & $-0.042^{+0.007}_{-0.007}$  \\ [0.9ex] 
 $\tau_U \,[\rm{Gyrs}]$  & $13.019^{+0.492}_{-0.389}$  & $15.148^{+1.745}_{-1.452}$  & $12.981^{+0.296}_{-0.292}$  & $13.564^{+1.133}_{-0.658}$  & $11.958^{+0.139}_{-0.109}$  & $12.601^{+0.067}_{-0.066}$  \\ [0.9ex] 
 $z_T $  & $0.718^{+0.087}_{-0.070}$  & $1.507^{+0.493}_{-1.407}$  & $0.685^{+0.049}_{-0.050}$  & $0.809^{+0.211}_{-0.124}$  & $0.652^{+0.032}_{-0.032}$  & $0.620^{+0.025}_{-0.025}$  \\ [0.9ex] 
 $q_0 $  & $-0.698^{+0.060}_{-0.042}$  & $-0.273^{+0.076}_{-0.071}$  & $-0.569^{+0.030}_{-0.028}$  & $-0.714^{+0.121}_{-0.059}$  & $-0.695^{+0.024}_{-0.020}$  & $-0.530^{+0.016}_{-0.016}$  \\ [0.9ex] 
$\chi^2$  & $15.67$  & $1989.52$  & $12.78$  & $435.88$  & $3115.60$  & $5627.80$  \\ [0.9ex] 

\hline
	\end{tabular}
\end{table*}

\begin{figure*}
    \centering
    \includegraphics[width=0.45\textwidth]{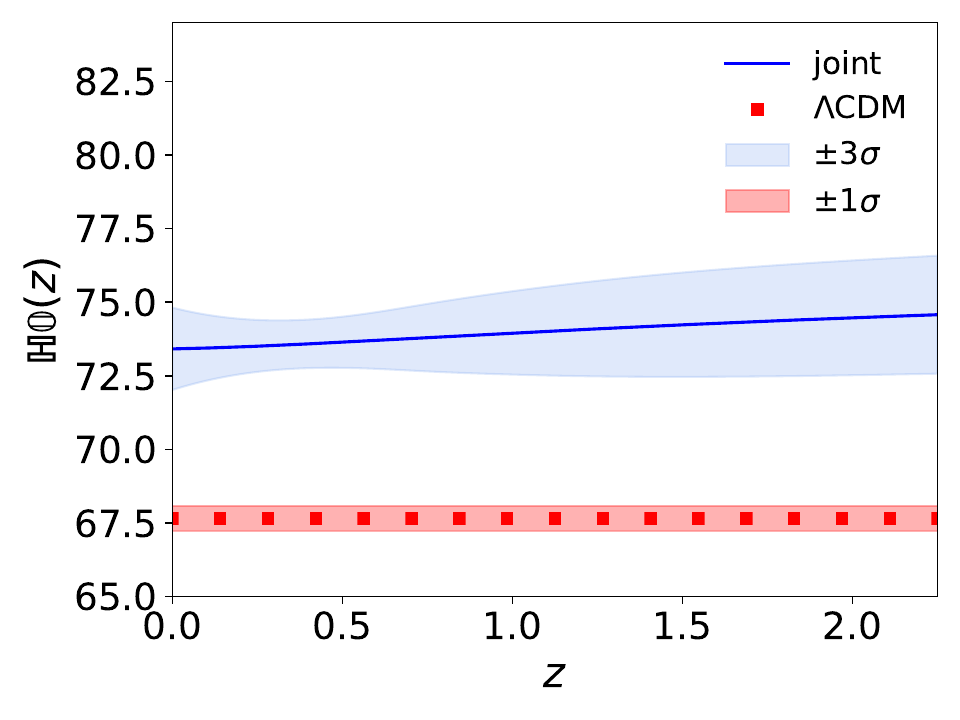}
    \includegraphics[width=0.45\textwidth]{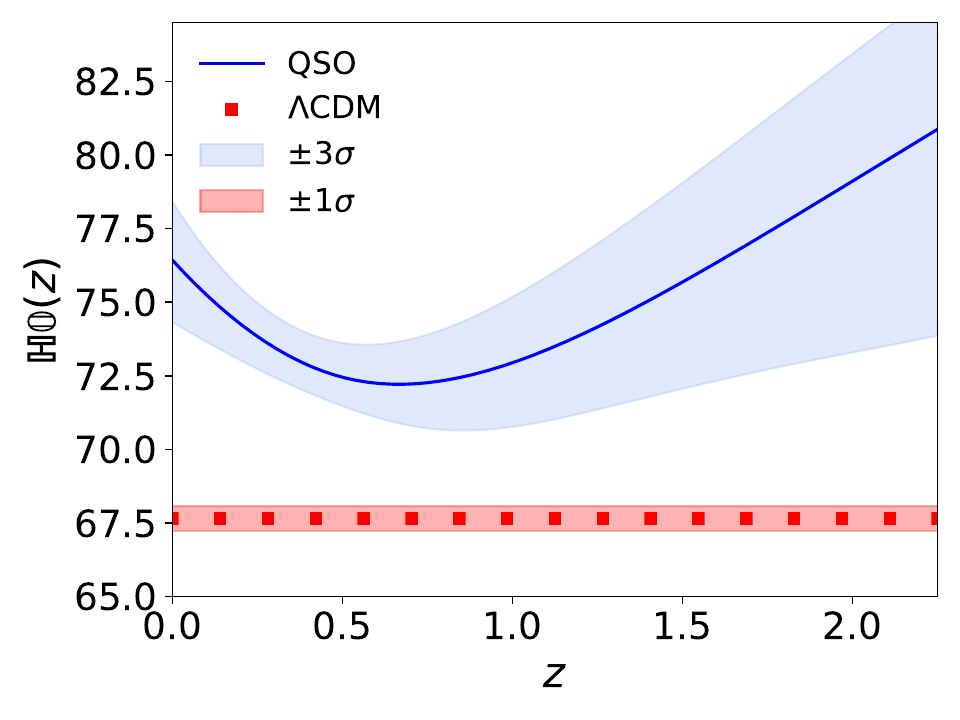}
    \caption{The $\mathbb{H}0(z)$ diagnostic for Rastall gravity, showing $\Lambda$CDM-R model. Here we present the behavior of $\mathbb{H}0(z)$ under the constraints of the QSO and joint analysis. The dotted red line represents the results for $\Lambda$CDM cosmology assuming $h=0.6766$ and $\Omega_{0m}=0.3111$ according to \cite{Planck:2018}.}
    \label{fig:H0diagnostic}
\end{figure*}

%%%%%%%%%%%%%%%%%%%%%%%%%%%%%%%%%%%%%%
\section{Conclusions and Outlooks} \label{CO}
%%%%%%%%%%%%%%%%%%%%%%%%%%%%%%%%%%%%%%

The objective of this paper is to study the performance of a recent datasample of QSO to constraint cosmological models and its interaction with other cosmological sample in a joint analysis. As a tester, we use Rastall gravity in the $\Lambda$CDM-R scenario, because it is demonstrated the equivalence with GR but with an extra free parameter, i.e. we have a theory in a gap between GR and a modification to GR.
 Results show that QSO sample presents a slight tension with other samples as it is observed in Fig. \ref{fig:contours}. ( The consistency with other samples is always maintained within $3\sigma$, even in the worst cases). Presenting a Universe's age younger than expected in general, additionally, the Rastall $\lambda$ parameter is dominant and negative suggesting a modification to GR. However, the performance of QSO in a joint analysis is better and slightly consistent with standard cosmology but pointing to an Universe age lower than expected. We observe from $j$ parameter in QSO reconstruction that the causative of the Universe acceleration is a dark energy instead of a CC, thus Rastall gravity is playing a role according to QSO.
Moreover, the $H_0$ tension remains unsolvable from $\mathbb{H}0(z)$ diagnostic when it is analyzed through the vision of a joint analysis and using Rastall gravity. A visible advantage of QSO is its high redshift, which has been, until now, an unexplored region which can help us understand the evolution of the early Universe and, if necessary, come up with an extension to the standard $\Lambda$CDM model. Through this study it is not possible to elucidate whether we are observing modifications to GR at intermediate redhifts or if QSO contains a bias that prefers $\lambda\neq0$. Recurrently, new datasamples are calibrated to a standard candle through the $\Lambda$CDM model, but this does not seem to be the case of QSO due to the modifications to GR tendency. We remark the study developed by  \cite{ShuoQSO:2017} in which the authors use QSO sample to constraint different DE models. In this case, they conclude non important differences with the standard $\Lambda$CDM model. However, in our study we are able to detect imprints of Rastall gravity in the results. We can not conclude if QSO is capable of detecting Rastall gravity because other DE theories are transparent for QSO \cite{ShuoQSO:2017}. We ought to keep in mind as well that we are adding a CC in order to have a late acceleration in Rastall gravity, which could induce a bias when it is analyzed through QSO. In summary, deeper exploration is necessary with more QSO points in order to elucidate what is the cause of the observed tendency for $\lambda$.

Additionally, we remark that the purpose behind the use of HIIG datasample is to implement a first test of comparison with QSO. This is because in the era of JW telescope we have galaxies at high redshifts \cite{Padmanabhan:2023esp} that can be used as a tester together with QSO. Our results are studied with and without HIIG datasample concluding that both constraints are compatible, presenting negligible differences.

We emphasize that Rastall gravity is not a feasible theory for understanding the nature of the current Universe's acceleration due to the necessity to add a dark energy. The parameter $\lambda$ has not the capability to produce an Universe acceleration on its own, thus, maintaining the cosmological constant and the Universe's acceleration as a mystery. Additionally, at background level it is impossible to alleviate the tension in $H_0$ and possibly others tensions such as $\sigma_8$, where $\Lambda$CDM's inferred value from CMB anisotropies is not consistent with those from weak gravitational lensing measurements \cite{DES:2021wwk,KiDS:2020suj, Heymans_2021}. However, a robust analysis is necessary using CMB perturbations due to the fact that with some samples we were able to distinguish between GR and Rastall cosmology.

As a final comment, we suggest a continuous development of the techniques that use QSO as a sample to constraint $\Lambda$CDM or theories that modify GR due to the promising capability of exploring high redshifts, something that can give us insights on the evolution of our Universe and the fluids that dwell within. Results given by QSO suggest a non negligible contribution of Rastall theory, thus, we suggest a rigorous study using cosmological perturbations theory in order to conclude the validity of Rastall theory and its application to cosmology.

\begin{acknowledgements}
We thank the anonymous referee for their thoughtful remarks and suggestions. J.A. Astorga-Moreno acknowledges CONAHCyT for the support in the postdoctoral fellowship at CINVESTAV, M\'exico. K. J aknowledges the support of the Universidad Iberoamericana. M.A.G-A acknowledges support from c\'atedra Marcos Moshinsky (MM) and Universidad Iberoamericana for the support with the National Research System (SNI) grant. The numerical analysis was carried out by the {\it Numerical Integration for Cosmological Theory and Experiments in High-energy Astrophysics} "Nicte Ha" cluster at IBERO University, acquired through c\'atedra MM support. A.H.A. thanks to the support from Luis Aguilar, 
Alejandro de Le\'on, Carlos Flores, and Jair Garc\'ia of the Laboratorio Nacional de Visualizaci\'on Cient\'ifica Avanzada. M.A.G.-A. and A.H.A acknowledge partial support from project ANID Vinculaci\'on Internacional FOVI220144. 
\end{acknowledgements}

\bibliography{main}

\begin{thebibliography}{51}%
\makeatletter
\providecommand \@ifxundefined [1]{%
 \@ifx{#1\undefined}
}%
\providecommand \@ifnum [1]{%
 \ifnum #1\expandafter \@firstoftwo
 \else \expandafter \@secondoftwo
 \fi
}%
\providecommand \@ifx [1]{%
 \ifx #1\expandafter \@firstoftwo
 \else \expandafter \@secondoftwo
 \fi
}%
\providecommand \natexlab [1]{#1}%
\providecommand \enquote  [1]{``#1''}%
\providecommand \bibnamefont  [1]{#1}%
\providecommand \bibfnamefont [1]{#1}%
\providecommand \citenamefont [1]{#1}%
\providecommand \href@noop [0]{\@secondoftwo}%
\providecommand \href [0]{\begingroup \@sanitize@url \@href}%
\providecommand \@href[1]{\@@startlink{#1}\@@href}%
\providecommand \@@href[1]{\endgroup#1\@@endlink}%
\providecommand \@sanitize@url [0]{\catcode `\\12\catcode `\$12\catcode
  `\&12\catcode `\#12\catcode `\^12\catcode `\_12\catcode `\%12\relax}%
\providecommand \@@startlink[1]{}%
\providecommand \@@endlink[0]{}%
\providecommand \url  [0]{\begingroup\@sanitize@url \@url }%
\providecommand \@url [1]{\endgroup\@href {#1}{\urlprefix }}%
\providecommand \urlprefix  [0]{URL }%
\providecommand \Eprint [0]{\href }%
\providecommand \doibase [0]{http://dx.doi.org/}%
\providecommand \selectlanguage [0]{\@gobble}%
\providecommand \bibinfo  [0]{\@secondoftwo}%
\providecommand \bibfield  [0]{\@secondoftwo}%
\providecommand \translation [1]{[#1]}%
\providecommand \BibitemOpen [0]{}%
\providecommand \bibitemStop [0]{}%
\providecommand \bibitemNoStop [0]{.\EOS\space}%
\providecommand \EOS [0]{\spacefactor3000\relax}%
\providecommand \BibitemShut  [1]{\csname bibitem#1\endcsname}%
\let\auto@bib@innerbib\@empty
%</preamble>
\bibitem [{\citenamefont {Perlmutter}\ \emph {et~al.}(1999)\citenamefont
  {Perlmutter}, \citenamefont {Aldering}, \citenamefont {Goldhaber},
  \citenamefont {Knop}, \citenamefont {Nugent}, \citenamefont {others},\ and\
  \citenamefont {Project}}]{Perlmutter:1999}%
  \BibitemOpen
  \bibfield  {author} {\bibinfo {author} {\bibfnamefont {S.}~\bibnamefont
  {Perlmutter}}, \bibinfo {author} {\bibfnamefont {G.}~\bibnamefont
  {Aldering}}, \bibinfo {author} {\bibfnamefont {G.}~\bibnamefont {Goldhaber}},
  \bibinfo {author} {\bibfnamefont {R.~A.}\ \bibnamefont {Knop}}, \bibinfo
  {author} {\bibfnamefont {P.}~\bibnamefont {Nugent}}, \bibinfo {author}
  {\bibnamefont {others}}, \ and\ \bibinfo {author} {\bibfnamefont {T.~S.~C.}\
  \bibnamefont {Project}},\ }\href
  {http://stacks.iop.org/0004-637X/517/i=2/a=565} {\bibfield  {journal}
  {\bibinfo  {journal} {The Astrophysical Journal}\ }\textbf {\bibinfo {volume}
  {517}},\ \bibinfo {pages} {565} (\bibinfo {year} {1999})}\BibitemShut
  {NoStop}%
\bibitem [{\citenamefont {Riess}\ \emph {et~al.}(1998)\citenamefont {Riess},
  \citenamefont {Filippenko}, \citenamefont {Challis}, \citenamefont
  {Clocchiatti}, \citenamefont {Diercks} \emph {et~al.}}]{Riess:1998}%
  \BibitemOpen
  \bibfield  {author} {\bibinfo {author} {\bibfnamefont {A.~G.}\ \bibnamefont
  {Riess}}, \bibinfo {author} {\bibfnamefont {A.~V.}\ \bibnamefont
  {Filippenko}}, \bibinfo {author} {\bibfnamefont {P.}~\bibnamefont {Challis}},
  \bibinfo {author} {\bibfnamefont {A.}~\bibnamefont {Clocchiatti}}, \bibinfo
  {author} {\bibfnamefont {A.}~\bibnamefont {Diercks}},  \emph {et~al.},\
  }\href {http://stacks.iop.org/1538-3881/116/i=3/a=1009} {\bibfield  {journal}
  {\bibinfo  {journal} {The Astronomical Journal}\ }\textbf {\bibinfo {volume}
  {116}},\ \bibinfo {pages} {1009} (\bibinfo {year} {1998})}\BibitemShut
  {NoStop}%
\bibitem [{\citenamefont {Aghanim}\ \emph {et~al.}(2020)\citenamefont
  {Aghanim}, \citenamefont {Akrami}, \citenamefont {Ashdown}, \citenamefont
  {Aumont}, \citenamefont {Baccigalupi}, \citenamefont {Ballardini},
  \citenamefont {Banday}, \citenamefont {Barreiro}, \citenamefont {Bartolo},
  \citenamefont {Basak}, \citenamefont {Battye}, \citenamefont {Benabed},
  \citenamefont {Bernard}, \citenamefont {Bersanelli}, \citenamefont
  {Bielewicz}, \citenamefont {Bock}, \citenamefont {Bond}, \citenamefont
  {Borrill}, \citenamefont {Bouchet}, \citenamefont {Boulanger}, \citenamefont
  {Bucher}, \citenamefont {Burigana}, \citenamefont {Butler}, \citenamefont
  {Calabrese}, \citenamefont {Cardoso}, \citenamefont {Carron}, \citenamefont
  {Challinor}, \citenamefont {Chiang}, \citenamefont {Chluba}, \citenamefont
  {Colombo}, \citenamefont {Combet}, \citenamefont {Contreras}, \citenamefont
  {Crill}, \citenamefont {Cuttaia}, \citenamefont {de~Bernardis}, \citenamefont
  {de~Zotti}, \citenamefont {Delabrouille}, \citenamefont {Delouis},
  \citenamefont {Valentino}, \citenamefont {Diego}, \citenamefont {Dor{\'{e}
  }}, \citenamefont {Douspis}, \citenamefont {Ducout}, \citenamefont {Dupac},
  \citenamefont {Dusini}, \citenamefont {Efstathiou}, \citenamefont {Elsner},
  \citenamefont {En{\ss}lin}, \citenamefont {Eriksen}, \citenamefont {Fantaye},
  \citenamefont {Farhang}, \citenamefont {Fergusson}, \citenamefont
  {Fernandez-Cobos}, \citenamefont {Finelli}, \citenamefont {Forastieri},
  \citenamefont {Frailis}, \citenamefont {Fraisse}, \citenamefont {Franceschi},
  \citenamefont {Frolov}, \citenamefont {Galeotta}, \citenamefont {Galli},
  \citenamefont {Ganga}, \citenamefont {G{\'{e}}nova-Santos}, \citenamefont
  {Gerbino}, \citenamefont {Ghosh}, \citenamefont {Gonz{\'{a}}lez-Nuevo},
  \citenamefont {G{\'{o}}rski}, \citenamefont {Gratton}, \citenamefont
  {Gruppuso}, \citenamefont {Gudmundsson}, \citenamefont {Hamann},
  \citenamefont {Handley}, \citenamefont {Hansen}, \citenamefont {Herranz},
  \citenamefont {Hildebrandt}, \citenamefont {Hivon}, \citenamefont {Huang},
  \citenamefont {Jaffe}, \citenamefont {Jones}, \citenamefont {Karakci},
  \citenamefont {Keihänen}, \citenamefont {Keskitalo}, \citenamefont
  {Kiiveri}, \citenamefont {Kim}, \citenamefont {Kisner}, \citenamefont {Knox},
  \citenamefont {Krachmalnicoff}, \citenamefont {Kunz}, \citenamefont
  {Kurki-Suonio}, \citenamefont {Lagache}, \citenamefont {Lamarre},
  \citenamefont {Lasenby}, \citenamefont {Lattanzi}, \citenamefont {Lawrence},
  \citenamefont {Jeune}, \citenamefont {Lemos}, \citenamefont {Lesgourgues},
  \citenamefont {Levrier}, \citenamefont {Lewis}, \citenamefont {Liguori},
  \citenamefont {Lilje}, \citenamefont {Lilley}, \citenamefont {Lindholm},
  \citenamefont {L{\'{o}}pez-Caniego}, \citenamefont {Lubin}, \citenamefont
  {Ma}, \citenamefont {Mac{\'{\i}}as-P{\'{e}}rez}, \citenamefont {Maggio},
  \citenamefont {Maino}, \citenamefont {Mandolesi}, \citenamefont {Mangilli},
  \citenamefont {Marcos-Caballero}, \citenamefont {Maris}, \citenamefont
  {Martin}, \citenamefont {Martinelli}, \citenamefont
  {Mart{\'{\i}}nez-Gonz{\'{a}}lez}, \citenamefont {Matarrese}, \citenamefont
  {Mauri}, \citenamefont {McEwen}, \citenamefont {Meinhold}, \citenamefont
  {Melchiorri}, \citenamefont {Mennella}, \citenamefont {Migliaccio},
  \citenamefont {Millea}, \citenamefont {Mitra}, \citenamefont
  {Miville-Desch{\^{e}}nes}, \citenamefont {Molinari}, \citenamefont {Montier},
  \citenamefont {Morgante}, \citenamefont {Moss}, \citenamefont {Natoli},
  \citenamefont {N{\o}rgaard-Nielsen}, \citenamefont {Pagano}, \citenamefont
  {Paoletti}, \citenamefont {Partridge}, \citenamefont {Patanchon},
  \citenamefont {Peiris}, \citenamefont {Perrotta}, \citenamefont {Pettorino},
  \citenamefont {Piacentini}, \citenamefont {Polastri}, \citenamefont
  {Polenta}, \citenamefont {Puget}, \citenamefont {Rachen}, \citenamefont
  {Reinecke}, \citenamefont {Remazeilles}, \citenamefont {Renzi}, \citenamefont
  {Rocha}, \citenamefont {Rosset}, \citenamefont {Roudier}, \citenamefont
  {Rubi{\~{n}}o-Mart{\'{\i}}n}, \citenamefont {Ruiz-Granados}, \citenamefont
  {Salvati}, \citenamefont {Sandri}, \citenamefont {Savelainen}, \citenamefont
  {Scott}, \citenamefont {Shellard}, \citenamefont {Sirignano}, \citenamefont
  {Sirri}, \citenamefont {Spencer}, \citenamefont {Sunyaev}, \citenamefont
  {Suur-Uski}, \citenamefont {Tauber}, \citenamefont {Tavagnacco},
  \citenamefont {Tenti}, \citenamefont {Toffolatti}, \citenamefont {Tomasi},
  \citenamefont {Trombetti}, \citenamefont {Valenziano}, \citenamefont
  {Valiviita}, \citenamefont {Tent}, \citenamefont {Vibert}, \citenamefont
  {Vielva}, \citenamefont {Villa}, \citenamefont {Vittorio}, \citenamefont
  {Wandelt}, \citenamefont {Wehus}, \citenamefont {White}, \citenamefont
  {White}, \citenamefont {Zacchei},\ and\ \citenamefont {Zonca}}]{Planck:2018}%
  \BibitemOpen
  \bibfield  {author} {\bibinfo {author} {\bibfnamefont {N.}~\bibnamefont
  {Aghanim}}, \bibinfo {author} {\bibfnamefont {Y.}~\bibnamefont {Akrami}},
  \bibinfo {author} {\bibfnamefont {M.}~\bibnamefont {Ashdown}}, \bibinfo
  {author} {\bibfnamefont {J.}~\bibnamefont {Aumont}}, \bibinfo {author}
  {\bibfnamefont {C.}~\bibnamefont {Baccigalupi}}, \bibinfo {author}
  {\bibfnamefont {M.}~\bibnamefont {Ballardini}}, \bibinfo {author}
  {\bibfnamefont {A.~J.}\ \bibnamefont {Banday}}, \bibinfo {author}
  {\bibfnamefont {R.~B.}\ \bibnamefont {Barreiro}}, \bibinfo {author}
  {\bibfnamefont {N.}~\bibnamefont {Bartolo}}, \bibinfo {author} {\bibfnamefont
  {S.}~\bibnamefont {Basak}}, \bibinfo {author} {\bibfnamefont
  {R.}~\bibnamefont {Battye}}, \bibinfo {author} {\bibfnamefont
  {K.}~\bibnamefont {Benabed}}, \bibinfo {author} {\bibfnamefont {J.-P.}\
  \bibnamefont {Bernard}}, \bibinfo {author} {\bibfnamefont {M.}~\bibnamefont
  {Bersanelli}}, \bibinfo {author} {\bibfnamefont {P.}~\bibnamefont
  {Bielewicz}}, \bibinfo {author} {\bibfnamefont {J.~J.}\ \bibnamefont {Bock}},
  \bibinfo {author} {\bibfnamefont {J.~R.}\ \bibnamefont {Bond}}, \bibinfo
  {author} {\bibfnamefont {J.}~\bibnamefont {Borrill}}, \bibinfo {author}
  {\bibfnamefont {F.~R.}\ \bibnamefont {Bouchet}}, \bibinfo {author}
  {\bibfnamefont {F.}~\bibnamefont {Boulanger}}, \bibinfo {author}
  {\bibfnamefont {M.}~\bibnamefont {Bucher}}, \bibinfo {author} {\bibfnamefont
  {C.}~\bibnamefont {Burigana}}, \bibinfo {author} {\bibfnamefont {R.~C.}\
  \bibnamefont {Butler}}, \bibinfo {author} {\bibfnamefont {E.}~\bibnamefont
  {Calabrese}}, \bibinfo {author} {\bibfnamefont {J.-F.}\ \bibnamefont
  {Cardoso}}, \bibinfo {author} {\bibfnamefont {J.}~\bibnamefont {Carron}},
  \bibinfo {author} {\bibfnamefont {A.}~\bibnamefont {Challinor}}, \bibinfo
  {author} {\bibfnamefont {H.~C.}\ \bibnamefont {Chiang}}, \bibinfo {author}
  {\bibfnamefont {J.}~\bibnamefont {Chluba}}, \bibinfo {author} {\bibfnamefont
  {L.~P.~L.}\ \bibnamefont {Colombo}}, \bibinfo {author} {\bibfnamefont
  {C.}~\bibnamefont {Combet}}, \bibinfo {author} {\bibfnamefont
  {D.}~\bibnamefont {Contreras}}, \bibinfo {author} {\bibfnamefont {B.~P.}\
  \bibnamefont {Crill}}, \bibinfo {author} {\bibfnamefont {F.}~\bibnamefont
  {Cuttaia}}, \bibinfo {author} {\bibfnamefont {P.}~\bibnamefont
  {de~Bernardis}}, \bibinfo {author} {\bibfnamefont {G.}~\bibnamefont
  {de~Zotti}}, \bibinfo {author} {\bibfnamefont {J.}~\bibnamefont
  {Delabrouille}}, \bibinfo {author} {\bibfnamefont {J.-M.}\ \bibnamefont
  {Delouis}}, \bibinfo {author} {\bibfnamefont {E.~D.}\ \bibnamefont
  {Valentino}}, \bibinfo {author} {\bibfnamefont {J.~M.}\ \bibnamefont
  {Diego}}, \bibinfo {author} {\bibfnamefont {O.}~\bibnamefont {Dor{\'{e} }}},
  \bibinfo {author} {\bibfnamefont {M.}~\bibnamefont {Douspis}}, \bibinfo
  {author} {\bibfnamefont {A.}~\bibnamefont {Ducout}}, \bibinfo {author}
  {\bibfnamefont {X.}~\bibnamefont {Dupac}}, \bibinfo {author} {\bibfnamefont
  {S.}~\bibnamefont {Dusini}}, \bibinfo {author} {\bibfnamefont
  {G.}~\bibnamefont {Efstathiou}}, \bibinfo {author} {\bibfnamefont
  {F.}~\bibnamefont {Elsner}}, \bibinfo {author} {\bibfnamefont {T.~A.}\
  \bibnamefont {En{\ss}lin}}, \bibinfo {author} {\bibfnamefont {H.~K.}\
  \bibnamefont {Eriksen}}, \bibinfo {author} {\bibfnamefont {Y.}~\bibnamefont
  {Fantaye}}, \bibinfo {author} {\bibfnamefont {M.}~\bibnamefont {Farhang}},
  \bibinfo {author} {\bibfnamefont {J.}~\bibnamefont {Fergusson}}, \bibinfo
  {author} {\bibfnamefont {R.}~\bibnamefont {Fernandez-Cobos}}, \bibinfo
  {author} {\bibfnamefont {F.}~\bibnamefont {Finelli}}, \bibinfo {author}
  {\bibfnamefont {F.}~\bibnamefont {Forastieri}}, \bibinfo {author}
  {\bibfnamefont {M.}~\bibnamefont {Frailis}}, \bibinfo {author} {\bibfnamefont
  {A.~A.}\ \bibnamefont {Fraisse}}, \bibinfo {author} {\bibfnamefont
  {E.}~\bibnamefont {Franceschi}}, \bibinfo {author} {\bibfnamefont
  {A.}~\bibnamefont {Frolov}}, \bibinfo {author} {\bibfnamefont
  {S.}~\bibnamefont {Galeotta}}, \bibinfo {author} {\bibfnamefont
  {S.}~\bibnamefont {Galli}}, \bibinfo {author} {\bibfnamefont
  {K.}~\bibnamefont {Ganga}}, \bibinfo {author} {\bibfnamefont {R.~T.}\
  \bibnamefont {G{\'{e}}nova-Santos}}, \bibinfo {author} {\bibfnamefont
  {M.}~\bibnamefont {Gerbino}}, \bibinfo {author} {\bibfnamefont
  {T.}~\bibnamefont {Ghosh}}, \bibinfo {author} {\bibfnamefont
  {J.}~\bibnamefont {Gonz{\'{a}}lez-Nuevo}}, \bibinfo {author} {\bibfnamefont
  {K.~M.}\ \bibnamefont {G{\'{o}}rski}}, \bibinfo {author} {\bibfnamefont
  {S.}~\bibnamefont {Gratton}}, \bibinfo {author} {\bibfnamefont
  {A.}~\bibnamefont {Gruppuso}}, \bibinfo {author} {\bibfnamefont {J.~E.}\
  \bibnamefont {Gudmundsson}}, \bibinfo {author} {\bibfnamefont
  {J.}~\bibnamefont {Hamann}}, \bibinfo {author} {\bibfnamefont
  {W.}~\bibnamefont {Handley}}, \bibinfo {author} {\bibfnamefont {F.~K.}\
  \bibnamefont {Hansen}}, \bibinfo {author} {\bibfnamefont {D.}~\bibnamefont
  {Herranz}}, \bibinfo {author} {\bibfnamefont {S.~R.}\ \bibnamefont
  {Hildebrandt}}, \bibinfo {author} {\bibfnamefont {E.}~\bibnamefont {Hivon}},
  \bibinfo {author} {\bibfnamefont {Z.}~\bibnamefont {Huang}}, \bibinfo
  {author} {\bibfnamefont {A.~H.}\ \bibnamefont {Jaffe}}, \bibinfo {author}
  {\bibfnamefont {W.~C.}\ \bibnamefont {Jones}}, \bibinfo {author}
  {\bibfnamefont {A.}~\bibnamefont {Karakci}}, \bibinfo {author} {\bibfnamefont
  {E.}~\bibnamefont {Keihänen}}, \bibinfo {author} {\bibfnamefont
  {R.}~\bibnamefont {Keskitalo}}, \bibinfo {author} {\bibfnamefont
  {K.}~\bibnamefont {Kiiveri}}, \bibinfo {author} {\bibfnamefont
  {J.}~\bibnamefont {Kim}}, \bibinfo {author} {\bibfnamefont {T.~S.}\
  \bibnamefont {Kisner}}, \bibinfo {author} {\bibfnamefont {L.}~\bibnamefont
  {Knox}}, \bibinfo {author} {\bibfnamefont {N.}~\bibnamefont
  {Krachmalnicoff}}, \bibinfo {author} {\bibfnamefont {M.}~\bibnamefont
  {Kunz}}, \bibinfo {author} {\bibfnamefont {H.}~\bibnamefont {Kurki-Suonio}},
  \bibinfo {author} {\bibfnamefont {G.}~\bibnamefont {Lagache}}, \bibinfo
  {author} {\bibfnamefont {J.-M.}\ \bibnamefont {Lamarre}}, \bibinfo {author}
  {\bibfnamefont {A.}~\bibnamefont {Lasenby}}, \bibinfo {author} {\bibfnamefont
  {M.}~\bibnamefont {Lattanzi}}, \bibinfo {author} {\bibfnamefont {C.~R.}\
  \bibnamefont {Lawrence}}, \bibinfo {author} {\bibfnamefont {M.~L.}\
  \bibnamefont {Jeune}}, \bibinfo {author} {\bibfnamefont {P.}~\bibnamefont
  {Lemos}}, \bibinfo {author} {\bibfnamefont {J.}~\bibnamefont {Lesgourgues}},
  \bibinfo {author} {\bibfnamefont {F.}~\bibnamefont {Levrier}}, \bibinfo
  {author} {\bibfnamefont {A.}~\bibnamefont {Lewis}}, \bibinfo {author}
  {\bibfnamefont {M.}~\bibnamefont {Liguori}}, \bibinfo {author} {\bibfnamefont
  {P.~B.}\ \bibnamefont {Lilje}}, \bibinfo {author} {\bibfnamefont
  {M.}~\bibnamefont {Lilley}}, \bibinfo {author} {\bibfnamefont
  {V.}~\bibnamefont {Lindholm}}, \bibinfo {author} {\bibfnamefont
  {M.}~\bibnamefont {L{\'{o}}pez-Caniego}}, \bibinfo {author} {\bibfnamefont
  {P.~M.}\ \bibnamefont {Lubin}}, \bibinfo {author} {\bibfnamefont {Y.-Z.}\
  \bibnamefont {Ma}}, \bibinfo {author} {\bibfnamefont {J.~F.}\ \bibnamefont
  {Mac{\'{\i}}as-P{\'{e}}rez}}, \bibinfo {author} {\bibfnamefont
  {G.}~\bibnamefont {Maggio}}, \bibinfo {author} {\bibfnamefont
  {D.}~\bibnamefont {Maino}}, \bibinfo {author} {\bibfnamefont
  {N.}~\bibnamefont {Mandolesi}}, \bibinfo {author} {\bibfnamefont
  {A.}~\bibnamefont {Mangilli}}, \bibinfo {author} {\bibfnamefont
  {A.}~\bibnamefont {Marcos-Caballero}}, \bibinfo {author} {\bibfnamefont
  {M.}~\bibnamefont {Maris}}, \bibinfo {author} {\bibfnamefont {P.~G.}\
  \bibnamefont {Martin}}, \bibinfo {author} {\bibfnamefont {M.}~\bibnamefont
  {Martinelli}}, \bibinfo {author} {\bibfnamefont {E.}~\bibnamefont
  {Mart{\'{\i}}nez-Gonz{\'{a}}lez}}, \bibinfo {author} {\bibfnamefont
  {S.}~\bibnamefont {Matarrese}}, \bibinfo {author} {\bibfnamefont
  {N.}~\bibnamefont {Mauri}}, \bibinfo {author} {\bibfnamefont {J.~D.}\
  \bibnamefont {McEwen}}, \bibinfo {author} {\bibfnamefont {P.~R.}\
  \bibnamefont {Meinhold}}, \bibinfo {author} {\bibfnamefont {A.}~\bibnamefont
  {Melchiorri}}, \bibinfo {author} {\bibfnamefont {A.}~\bibnamefont
  {Mennella}}, \bibinfo {author} {\bibfnamefont {M.}~\bibnamefont
  {Migliaccio}}, \bibinfo {author} {\bibfnamefont {M.}~\bibnamefont {Millea}},
  \bibinfo {author} {\bibfnamefont {S.}~\bibnamefont {Mitra}}, \bibinfo
  {author} {\bibfnamefont {M.-A.}\ \bibnamefont {Miville-Desch{\^{e}}nes}},
  \bibinfo {author} {\bibfnamefont {D.}~\bibnamefont {Molinari}}, \bibinfo
  {author} {\bibfnamefont {L.}~\bibnamefont {Montier}}, \bibinfo {author}
  {\bibfnamefont {G.}~\bibnamefont {Morgante}}, \bibinfo {author}
  {\bibfnamefont {A.}~\bibnamefont {Moss}}, \bibinfo {author} {\bibfnamefont
  {P.}~\bibnamefont {Natoli}}, \bibinfo {author} {\bibfnamefont {H.~U.}\
  \bibnamefont {N{\o}rgaard-Nielsen}}, \bibinfo {author} {\bibfnamefont
  {L.}~\bibnamefont {Pagano}}, \bibinfo {author} {\bibfnamefont
  {D.}~\bibnamefont {Paoletti}}, \bibinfo {author} {\bibfnamefont
  {B.}~\bibnamefont {Partridge}}, \bibinfo {author} {\bibfnamefont
  {G.}~\bibnamefont {Patanchon}}, \bibinfo {author} {\bibfnamefont {H.~V.}\
  \bibnamefont {Peiris}}, \bibinfo {author} {\bibfnamefont {F.}~\bibnamefont
  {Perrotta}}, \bibinfo {author} {\bibfnamefont {V.}~\bibnamefont {Pettorino}},
  \bibinfo {author} {\bibfnamefont {F.}~\bibnamefont {Piacentini}}, \bibinfo
  {author} {\bibfnamefont {L.}~\bibnamefont {Polastri}}, \bibinfo {author}
  {\bibfnamefont {G.}~\bibnamefont {Polenta}}, \bibinfo {author} {\bibfnamefont
  {J.-L.}\ \bibnamefont {Puget}}, \bibinfo {author} {\bibfnamefont {J.~P.}\
  \bibnamefont {Rachen}}, \bibinfo {author} {\bibfnamefont {M.}~\bibnamefont
  {Reinecke}}, \bibinfo {author} {\bibfnamefont {M.}~\bibnamefont
  {Remazeilles}}, \bibinfo {author} {\bibfnamefont {A.}~\bibnamefont {Renzi}},
  \bibinfo {author} {\bibfnamefont {G.}~\bibnamefont {Rocha}}, \bibinfo
  {author} {\bibfnamefont {C.}~\bibnamefont {Rosset}}, \bibinfo {author}
  {\bibfnamefont {G.}~\bibnamefont {Roudier}}, \bibinfo {author} {\bibfnamefont
  {J.~A.}\ \bibnamefont {Rubi{\~{n}}o-Mart{\'{\i}}n}}, \bibinfo {author}
  {\bibfnamefont {B.}~\bibnamefont {Ruiz-Granados}}, \bibinfo {author}
  {\bibfnamefont {L.}~\bibnamefont {Salvati}}, \bibinfo {author} {\bibfnamefont
  {M.}~\bibnamefont {Sandri}}, \bibinfo {author} {\bibfnamefont
  {M.}~\bibnamefont {Savelainen}}, \bibinfo {author} {\bibfnamefont
  {D.}~\bibnamefont {Scott}}, \bibinfo {author} {\bibfnamefont {E.~P.~S.}\
  \bibnamefont {Shellard}}, \bibinfo {author} {\bibfnamefont {C.}~\bibnamefont
  {Sirignano}}, \bibinfo {author} {\bibfnamefont {G.}~\bibnamefont {Sirri}},
  \bibinfo {author} {\bibfnamefont {L.~D.}\ \bibnamefont {Spencer}}, \bibinfo
  {author} {\bibfnamefont {R.}~\bibnamefont {Sunyaev}}, \bibinfo {author}
  {\bibfnamefont {A.-S.}\ \bibnamefont {Suur-Uski}}, \bibinfo {author}
  {\bibfnamefont {J.~A.}\ \bibnamefont {Tauber}}, \bibinfo {author}
  {\bibfnamefont {D.}~\bibnamefont {Tavagnacco}}, \bibinfo {author}
  {\bibfnamefont {M.}~\bibnamefont {Tenti}}, \bibinfo {author} {\bibfnamefont
  {L.}~\bibnamefont {Toffolatti}}, \bibinfo {author} {\bibfnamefont
  {M.}~\bibnamefont {Tomasi}}, \bibinfo {author} {\bibfnamefont
  {T.}~\bibnamefont {Trombetti}}, \bibinfo {author} {\bibfnamefont
  {L.}~\bibnamefont {Valenziano}}, \bibinfo {author} {\bibfnamefont
  {J.}~\bibnamefont {Valiviita}}, \bibinfo {author} {\bibfnamefont {B.~V.}\
  \bibnamefont {Tent}}, \bibinfo {author} {\bibfnamefont {L.}~\bibnamefont
  {Vibert}}, \bibinfo {author} {\bibfnamefont {P.}~\bibnamefont {Vielva}},
  \bibinfo {author} {\bibfnamefont {F.}~\bibnamefont {Villa}}, \bibinfo
  {author} {\bibfnamefont {N.}~\bibnamefont {Vittorio}}, \bibinfo {author}
  {\bibfnamefont {B.~D.}\ \bibnamefont {Wandelt}}, \bibinfo {author}
  {\bibfnamefont {I.~K.}\ \bibnamefont {Wehus}}, \bibinfo {author}
  {\bibfnamefont {M.}~\bibnamefont {White}}, \bibinfo {author} {\bibfnamefont
  {S.~D.~M.}\ \bibnamefont {White}}, \bibinfo {author} {\bibfnamefont
  {A.}~\bibnamefont {Zacchei}}, \ and\ \bibinfo {author} {\bibfnamefont
  {A.}~\bibnamefont {Zonca}},\ }\href {\doibase 10.1051/0004-6361/201833910}
  {\bibfield  {journal} {\bibinfo  {journal} {Astronomy {\&} Astrophysics}\
  }\textbf {\bibinfo {volume} {641}},\ \bibinfo {pages} {A6} (\bibinfo {year}
  {2020})}\BibitemShut {NoStop}%
\bibitem [{\citenamefont {Di~Valentino}\ \emph {et~al.}(2021)\citenamefont
  {Di~Valentino}, \citenamefont {Mena}, \citenamefont {Pan}, \citenamefont
  {Visinelli}, \citenamefont {Yang}, \citenamefont {Melchiorri}, \citenamefont
  {Mota}, \citenamefont {Riess},\ and\ \citenamefont
  {Silk}}]{DiValentino:2021izs}%
  \BibitemOpen
  \bibfield  {author} {\bibinfo {author} {\bibfnamefont {E.}~\bibnamefont
  {Di~Valentino}}, \bibinfo {author} {\bibfnamefont {O.}~\bibnamefont {Mena}},
  \bibinfo {author} {\bibfnamefont {S.}~\bibnamefont {Pan}}, \bibinfo {author}
  {\bibfnamefont {L.}~\bibnamefont {Visinelli}}, \bibinfo {author}
  {\bibfnamefont {W.}~\bibnamefont {Yang}}, \bibinfo {author} {\bibfnamefont
  {A.}~\bibnamefont {Melchiorri}}, \bibinfo {author} {\bibfnamefont {D.~F.}\
  \bibnamefont {Mota}}, \bibinfo {author} {\bibfnamefont {A.~G.}\ \bibnamefont
  {Riess}}, \ and\ \bibinfo {author} {\bibfnamefont {J.}~\bibnamefont {Silk}},\
  }\href {\doibase 10.1088/1361-6382/ac086d} {\bibfield  {journal} {\bibinfo
  {journal} {Class. Quant. Grav.}\ }\textbf {\bibinfo {volume} {38}},\ \bibinfo
  {pages} {153001} (\bibinfo {year} {2021})},\ \Eprint
  {http://arxiv.org/abs/2103.01183} {arXiv:2103.01183 [astro-ph.CO]}
  \BibitemShut {NoStop}%
\bibitem [{\citenamefont {Motta}\ \emph {et~al.}(2021)\citenamefont {Motta},
  \citenamefont {Garc\'\i{}a-Aspeitia}, \citenamefont {Hern\'andez-Almada},
  \citenamefont {Maga\~na},\ and\ \citenamefont {Verdugo}}]{Motta:2021hvl}%
  \BibitemOpen
  \bibfield  {author} {\bibinfo {author} {\bibfnamefont {V.}~\bibnamefont
  {Motta}}, \bibinfo {author} {\bibfnamefont {M.~A.}\ \bibnamefont
  {Garc\'\i{}a-Aspeitia}}, \bibinfo {author} {\bibfnamefont {A.}~\bibnamefont
  {Hern\'andez-Almada}}, \bibinfo {author} {\bibfnamefont {J.}~\bibnamefont
  {Maga\~na}}, \ and\ \bibinfo {author} {\bibfnamefont {T.}~\bibnamefont
  {Verdugo}},\ }\href {\doibase 10.3390/universe7060163} {\bibfield  {journal}
  {\bibinfo  {journal} {Universe}\ }\textbf {\bibinfo {volume} {7}},\ \bibinfo
  {pages} {163} (\bibinfo {year} {2021})},\ \Eprint
  {http://arxiv.org/abs/2104.04642} {arXiv:2104.04642 [astro-ph.CO]}
  \BibitemShut {NoStop}%
\bibitem [{\citenamefont {Carroll}(2001)}]{Carroll:2000}%
  \BibitemOpen
  \bibfield  {author} {\bibinfo {author} {\bibfnamefont {S.~M.}\ \bibnamefont
  {Carroll}},\ }\href {\doibase 10.12942/lrr-2001-1} {\bibfield  {journal}
  {\bibinfo  {journal} {Living Rev. Rel.}\ }\textbf {\bibinfo {volume} {4}},\
  \bibinfo {pages} {1} (\bibinfo {year} {2001})},\ \Eprint
  {http://arxiv.org/abs/astro-ph/0004075} {arXiv:astro-ph/0004075 [astro-ph]}
  \BibitemShut {NoStop}%
%%CITATION = ASTRO-PH/0004075;%%
\bibitem [{\citenamefont {Riess}\ \emph {et~al.}(2022)\citenamefont {Riess},
  \citenamefont {Yuan}, \citenamefont {Macri}, \citenamefont {Scolnic},
  \citenamefont {Brout}, \citenamefont {Casertano}, \citenamefont {Jones},
  \citenamefont {Murakami}, \citenamefont {Anand}, \citenamefont {Breuval},
  \citenamefont {Brink}, \citenamefont {Filippenko}, \citenamefont {Hoffmann},
  \citenamefont {Jha}, \citenamefont {Kenworthy}, \citenamefont {Mackenty},
  \citenamefont {Stahl},\ and\ \citenamefont {Zheng}}]{Riess:2021jrx}%
  \BibitemOpen
  \bibfield  {author} {\bibinfo {author} {\bibfnamefont {A.~G.}\ \bibnamefont
  {Riess}}, \bibinfo {author} {\bibfnamefont {W.}~\bibnamefont {Yuan}},
  \bibinfo {author} {\bibfnamefont {L.~M.}\ \bibnamefont {Macri}}, \bibinfo
  {author} {\bibfnamefont {D.}~\bibnamefont {Scolnic}}, \bibinfo {author}
  {\bibfnamefont {D.}~\bibnamefont {Brout}}, \bibinfo {author} {\bibfnamefont
  {S.}~\bibnamefont {Casertano}}, \bibinfo {author} {\bibfnamefont {D.~O.}\
  \bibnamefont {Jones}}, \bibinfo {author} {\bibfnamefont {Y.}~\bibnamefont
  {Murakami}}, \bibinfo {author} {\bibfnamefont {G.~S.}\ \bibnamefont {Anand}},
  \bibinfo {author} {\bibfnamefont {L.}~\bibnamefont {Breuval}}, \bibinfo
  {author} {\bibfnamefont {T.~G.}\ \bibnamefont {Brink}}, \bibinfo {author}
  {\bibfnamefont {A.~V.}\ \bibnamefont {Filippenko}}, \bibinfo {author}
  {\bibfnamefont {S.}~\bibnamefont {Hoffmann}}, \bibinfo {author}
  {\bibfnamefont {S.~W.}\ \bibnamefont {Jha}}, \bibinfo {author} {\bibfnamefont
  {W.~D.}\ \bibnamefont {Kenworthy}}, \bibinfo {author} {\bibfnamefont
  {J.}~\bibnamefont {Mackenty}}, \bibinfo {author} {\bibfnamefont {B.~E.}\
  \bibnamefont {Stahl}}, \ and\ \bibinfo {author} {\bibfnamefont
  {W.}~\bibnamefont {Zheng}},\ }\href {\doibase 10.3847/2041-8213/ac5c5b}
  {\bibfield  {journal} {\bibinfo  {journal} {The Astrophysical Journal
  Letters}\ }\textbf {\bibinfo {volume} {934}},\ \bibinfo {pages} {L7}
  (\bibinfo {year} {2022})}\BibitemShut {NoStop}%
\bibitem [{\citenamefont {Efstathiou}(2021)}]{Efstathiou:2021ocp}%
  \BibitemOpen
  \bibfield  {author} {\bibinfo {author} {\bibfnamefont {G.}~\bibnamefont
  {Efstathiou}},\ }\href {\doibase 10.1093/mnras/stab1588} {\bibfield
  {journal} {\bibinfo  {journal} {Mon. Not. Roy. Astron. Soc.}\ }\textbf
  {\bibinfo {volume} {505}},\ \bibinfo {pages} {3866} (\bibinfo {year}
  {2021})},\ \Eprint {http://arxiv.org/abs/2103.08723} {arXiv:2103.08723
  [astro-ph.CO]} \BibitemShut {NoStop}%
\bibitem [{\citenamefont {Amendola}\ and\ \citenamefont
  {Tsujikawa}(2007)}]{fr}%
  \BibitemOpen
  \bibfield  {author} {\bibinfo {author} {\bibfnamefont {G.~R.}\ \bibnamefont
  {Amendola}, \bibfnamefont {L.}}\ and\ \bibinfo {author} {\bibfnamefont
  {S.}~\bibnamefont {Tsujikawa}},\ }\href@noop {} {\bibfield  {journal}
  {\bibinfo  {journal} {Living Rev. Relativity}\ }\textbf {\bibinfo {volume}
  {13}} (\bibinfo {year} {2007})}\BibitemShut {NoStop}%
\bibitem [{\citenamefont {De~Felice}\ and\ \citenamefont
  {Tsujikawa}(2010)}]{fr2}%
  \BibitemOpen
  \bibfield  {author} {\bibinfo {author} {\bibfnamefont {A.}~\bibnamefont
  {De~Felice}}\ and\ \bibinfo {author} {\bibfnamefont {S.}~\bibnamefont
  {Tsujikawa}},\ }\href@noop {} {\bibfield  {journal} {\bibinfo  {journal}
  {Living Rev. Relativity}\ }\textbf {\bibinfo {volume} {13}} (\bibinfo {year}
  {2010})}\BibitemShut {NoStop}%
\bibitem [{\citenamefont {Shahidi}(2021)}]{fr3}%
  \BibitemOpen
  \bibfield  {author} {\bibinfo {author} {\bibfnamefont {S.}~\bibnamefont
  {Shahidi}},\ }\href@noop {} {\  (\bibinfo {year} {2021})},\ \Eprint
  {http://arxiv.org/abs/2108.00423v1} {asXiv:2108.00423v1 [gr-qc]} \BibitemShut
  {NoStop}%
\bibitem [{\citenamefont {Ayuso}\ \emph {et~al.}(2021)\citenamefont {Ayuso},
  \citenamefont {Lazkoz},\ and\ \citenamefont {Salzano}}]{Ayuso:2020dcu}%
  \BibitemOpen
  \bibfield  {author} {\bibinfo {author} {\bibfnamefont {I.}~\bibnamefont
  {Ayuso}}, \bibinfo {author} {\bibfnamefont {R.}~\bibnamefont {Lazkoz}}, \
  and\ \bibinfo {author} {\bibfnamefont {V.}~\bibnamefont {Salzano}},\ }\href
  {\doibase 10.1103/PhysRevD.103.063505} {\bibfield  {journal} {\bibinfo
  {journal} {Phys. Rev. D}\ }\textbf {\bibinfo {volume} {103}},\ \bibinfo
  {pages} {063505} (\bibinfo {year} {2021})},\ \Eprint
  {http://arxiv.org/abs/2012.00046} {arXiv:2012.00046 [astro-ph.CO]}
  \BibitemShut {NoStop}%
\bibitem [{\citenamefont {Ayuso}\ \emph {et~al.}(2022)\citenamefont {Ayuso},
  \citenamefont {Lazkoz},\ and\ \citenamefont {Mimoso}}]{Ayuso:2021vtj}%
  \BibitemOpen
  \bibfield  {author} {\bibinfo {author} {\bibfnamefont {I.}~\bibnamefont
  {Ayuso}}, \bibinfo {author} {\bibfnamefont {R.}~\bibnamefont {Lazkoz}}, \
  and\ \bibinfo {author} {\bibfnamefont {J.~P.}\ \bibnamefont {Mimoso}},\
  }\href {\doibase 10.1103/PhysRevD.105.083534} {\bibfield  {journal} {\bibinfo
   {journal} {Phys. Rev. D}\ }\textbf {\bibinfo {volume} {105}},\ \bibinfo
  {pages} {083534} (\bibinfo {year} {2022})},\ \Eprint
  {http://arxiv.org/abs/2111.05061} {arXiv:2111.05061 [astro-ph.CO]}
  \BibitemShut {NoStop}%
\bibitem [{\citenamefont {Gohar}\ and\ \citenamefont
  {Salzano}(2021)}]{Gohar:2020bod}%
  \BibitemOpen
  \bibfield  {author} {\bibinfo {author} {\bibfnamefont {H.}~\bibnamefont
  {Gohar}}\ and\ \bibinfo {author} {\bibfnamefont {V.}~\bibnamefont
  {Salzano}},\ }\href {\doibase 10.1140/epjc/s10052-021-09086-9} {\bibfield
  {journal} {\bibinfo  {journal} {Eur. Phys. J. C}\ }\textbf {\bibinfo {volume}
  {81}},\ \bibinfo {pages} {338} (\bibinfo {year} {2021})},\ \Eprint
  {http://arxiv.org/abs/2008.09635} {arXiv:2008.09635 [gr-qc]} \BibitemShut
  {NoStop}%
\bibitem [{\citenamefont {Lazkoz}\ \emph {et~al.}(2019)\citenamefont {Lazkoz},
  \citenamefont {Lobo}, \citenamefont {Ortiz-Ba\~nos},\ and\ \citenamefont
  {Salzano}}]{Lazkoz:2019sjl}%
  \BibitemOpen
  \bibfield  {author} {\bibinfo {author} {\bibfnamefont {R.}~\bibnamefont
  {Lazkoz}}, \bibinfo {author} {\bibfnamefont {F.~S.~N.}\ \bibnamefont {Lobo}},
  \bibinfo {author} {\bibfnamefont {M.}~\bibnamefont {Ortiz-Ba\~nos}}, \ and\
  \bibinfo {author} {\bibfnamefont {V.}~\bibnamefont {Salzano}},\ }\href
  {\doibase 10.1103/PhysRevD.100.104027} {\bibfield  {journal} {\bibinfo
  {journal} {Phys. Rev. D}\ }\textbf {\bibinfo {volume} {100}},\ \bibinfo
  {pages} {104027} (\bibinfo {year} {2019})},\ \Eprint
  {http://arxiv.org/abs/1907.13219} {arXiv:1907.13219 [gr-qc]} \BibitemShut
  {NoStop}%
\bibitem [{\citenamefont {Rastall}(1972)}]{rast}%
  \BibitemOpen
  \bibfield  {author} {\bibinfo {author} {\bibfnamefont {P.}~\bibnamefont
  {Rastall}},\ }\href@noop {} {\bibfield  {journal} {\bibinfo  {journal} {Phys.
  Rev. D}\ }\textbf {\bibinfo {volume} {6}},\ \bibinfo {pages} {3357} (\bibinfo
  {year} {1972})}\BibitemShut {NoStop}%
\bibitem [{\citenamefont {Rastall}(1976)}]{rast2}%
  \BibitemOpen
  \bibfield  {author} {\bibinfo {author} {\bibfnamefont {P.}~\bibnamefont
  {Rastall}},\ }\href@noop {} {\bibfield  {journal} {\bibinfo  {journal} {Can.
  J. Phys.}\ }\textbf {\bibinfo {volume} {54}} (\bibinfo {year}
  {1976})}\BibitemShut {NoStop}%
\bibitem [{\citenamefont {Batista}\ \emph {et~al.}(2012)\citenamefont
  {Batista}, \citenamefont {Daouda}, \citenamefont {Fabris},\ and\
  \citenamefont {Rodrigues}}]{l1}%
  \BibitemOpen
  \bibfield  {author} {\bibinfo {author} {\bibfnamefont {C.~E.~M.}\
  \bibnamefont {Batista}}, \bibinfo {author} {\bibfnamefont {M.~H.}\
  \bibnamefont {Daouda}}, \bibinfo {author} {\bibfnamefont {O.~F.}\
  \bibnamefont {Fabris}, \bibfnamefont {J.~C.and~Piatella}}, \ and\ \bibinfo
  {author} {\bibfnamefont {D.~C.}\ \bibnamefont {Rodrigues}},\ }\href@noop {}
  {\bibfield  {journal} {\bibinfo  {journal} {Phys. Rev.D}\ }\textbf {\bibinfo
  {volume} {85}},\ \bibinfo {pages} {084008} (\bibinfo {year}
  {2012})}\BibitemShut {NoStop}%
\bibitem [{\citenamefont {Batista}\ \emph {et~al.}(2013)\citenamefont
  {Batista}, \citenamefont {Fabris}, \citenamefont {Piatella},\ and\
  \citenamefont {M.}}]{l2}%
  \BibitemOpen
  \bibfield  {author} {\bibinfo {author} {\bibfnamefont {C.~E.~M.}\
  \bibnamefont {Batista}}, \bibinfo {author} {\bibfnamefont {J.~C.}\
  \bibnamefont {Fabris}}, \bibinfo {author} {\bibfnamefont {O.~F.}\
  \bibnamefont {Piatella}}, \ and\ \bibinfo {author} {\bibfnamefont {V.-T.~A.}\
  \bibnamefont {M.}},\ }\href@noop {} {\bibfield  {journal} {\bibinfo
  {journal} {Eur. Phys. J.}\ }\textbf {\bibinfo {volume} {C 73}},\ \bibinfo
  {pages} {2425} (\bibinfo {year} {2013})}\BibitemShut {NoStop}%
\bibitem [{\citenamefont {Van~de Bruck}\ and\ \citenamefont
  {Thomas}(2019)}]{l3}%
  \BibitemOpen
  \bibfield  {author} {\bibinfo {author} {\bibfnamefont {C.}~\bibnamefont
  {Van~de Bruck}}\ and\ \bibinfo {author} {\bibfnamefont {C.~C.}\ \bibnamefont
  {Thomas}},\ }\href@noop {} {\bibfield  {journal} {\bibinfo  {journal} {Phys.
  Rev. D}\ }\textbf {\bibinfo {volume} {100}},\ \bibinfo {pages} {023515}
  (\bibinfo {year} {2019})}\BibitemShut {NoStop}%
\bibitem [{\citenamefont {Akarsu}\ \emph {et~al.}(2020)\citenamefont {Akarsu},
  \citenamefont {Katirci}, \citenamefont {Kumar}, \citenamefont {Nunes},
  \citenamefont {{\"O}zt{\"u}rk},\ and\ \citenamefont {Sharma}}]{l4}%
  \BibitemOpen
  \bibfield  {author} {\bibinfo {author} {\bibfnamefont {{\"O}.}~\bibnamefont
  {Akarsu}}, \bibinfo {author} {\bibfnamefont {N.}~\bibnamefont {Katirci}},
  \bibinfo {author} {\bibfnamefont {S.}~\bibnamefont {Kumar}}, \bibinfo
  {author} {\bibfnamefont {R.}~\bibnamefont {Nunes}}, \bibinfo {author}
  {\bibfnamefont {B.}~\bibnamefont {{\"O}zt{\"u}rk}}, \ and\ \bibinfo {author}
  {\bibfnamefont {S.}~\bibnamefont {Sharma}},\ }\href@noop {} {\  (\bibinfo
  {year} {2020})},\ \Eprint {http://arxiv.org/abs/2004.04074v3}
  {arXiv:2004.04074v3 [astro-ph.CO]} \BibitemShut {NoStop}%
\bibitem [{\citenamefont {Oliveira}\ \emph {et~al.}(2015)\citenamefont
  {Oliveira}, \citenamefont {Velten}, \citenamefont {Fabris},\ and\
  \citenamefont {Casarini}}]{a}%
  \BibitemOpen
  \bibfield  {author} {\bibinfo {author} {\bibfnamefont {A.~M.}\ \bibnamefont
  {Oliveira}}, \bibinfo {author} {\bibfnamefont {H.~E.~S.}\ \bibnamefont
  {Velten}}, \bibinfo {author} {\bibfnamefont {J.~C.}\ \bibnamefont {Fabris}},
  \ and\ \bibinfo {author} {\bibfnamefont {L.}~\bibnamefont {Casarini}},\
  }\href@noop {} {\bibfield  {journal} {\bibinfo  {journal} {Phys. Rev. D}\
  }\textbf {\bibinfo {volume} {92}},\ \bibinfo {pages} {04420} (\bibinfo {year}
  {2015})}\BibitemShut {NoStop}%
\bibitem [{\citenamefont {Daouda}\ \emph {et~al.}(2019)\citenamefont {Daouda},
  \citenamefont {Fabris}, \citenamefont {Oliveira}, \citenamefont {Smirnov},\
  and\ \citenamefont {Velten}}]{b}%
  \BibitemOpen
  \bibfield  {author} {\bibinfo {author} {\bibfnamefont {M.}~\bibnamefont
  {Daouda}}, \bibinfo {author} {\bibfnamefont {J.~C.}\ \bibnamefont {Fabris}},
  \bibinfo {author} {\bibfnamefont {A.~M.}\ \bibnamefont {Oliveira}}, \bibinfo
  {author} {\bibfnamefont {F.}~\bibnamefont {Smirnov}}, \ and\ \bibinfo
  {author} {\bibfnamefont {H.~E.~S.}\ \bibnamefont {Velten}},\ }\href@noop {}
  {\bibfield  {journal} {\bibinfo  {journal} {Int. J. Mod Phys. D}\ }\textbf
  {\bibinfo {volume} {28}},\ \bibinfo {pages} {1950175} (\bibinfo {year}
  {2019})}\BibitemShut {NoStop}%
\bibitem [{\citenamefont {Hadi~Ziaie}\ \emph {et~al.}(2019)\citenamefont
  {Hadi~Ziaie}, \citenamefont {Moradpour},\ and\ \citenamefont {Ghaffari}}]{c}%
  \BibitemOpen
  \bibfield  {author} {\bibinfo {author} {\bibfnamefont {A.}~\bibnamefont
  {Hadi~Ziaie}}, \bibinfo {author} {\bibfnamefont {H.}~\bibnamefont
  {Moradpour}}, \ and\ \bibinfo {author} {\bibfnamefont {S.}~\bibnamefont
  {Ghaffari}},\ }\href@noop {} {\bibfield  {journal} {\bibinfo  {journal}
  {Phys. Lett.}\ }\textbf {\bibinfo {volume} {13}},\ \bibinfo {pages} {276}
  (\bibinfo {year} {2019})}\BibitemShut {NoStop}%
\bibitem [{\citenamefont {Shabani}\ and\ \citenamefont {Hadi~Ziaie}(2020)}]{d}%
  \BibitemOpen
  \bibfield  {author} {\bibinfo {author} {\bibfnamefont {H.}~\bibnamefont
  {Shabani}}\ and\ \bibinfo {author} {\bibfnamefont {A.}~\bibnamefont
  {Hadi~Ziaie}},\ }\href@noop {} {\bibfield  {journal} {\bibinfo  {journal}
  {EPL}\ }\textbf {\bibinfo {volume} {129}},\ \bibinfo {pages} {20004}
  (\bibinfo {year} {2020})}\BibitemShut {NoStop}%
\bibitem [{\citenamefont {Lusso}(2020)}]{Lusso:2020fax}%
  \BibitemOpen
  \bibfield  {author} {\bibinfo {author} {\bibfnamefont {E.}~\bibnamefont
  {Lusso}},\ }\href@noop {} {\  (\bibinfo {year} {2020})},\ \Eprint
  {http://arxiv.org/abs/2002.02464} {arXiv:2002.02464 [astro-ph.CO]}
  \BibitemShut {NoStop}%
\bibitem [{\citenamefont {{Peterson}}(2006)}]{peterson2004}%
  \BibitemOpen
  \bibfield  {author} {\bibinfo {author} {\bibfnamefont {B.~M.}\ \bibnamefont
  {{Peterson}}},\ }in\ \href {\doibase 10.1007/3-540-34621-X_3} {\emph
  {\bibinfo {booktitle} {Physics of Active Galactic Nuclei at all Scales}}},\
  Vol.\ \bibinfo {volume} {693},\ \bibinfo {editor} {edited by\ \bibinfo
  {editor} {\bibfnamefont {D.}~\bibnamefont {{Alloin}}}}\ (\bibinfo {year}
  {2006})\ p.~\bibinfo {pages} {77}\BibitemShut {NoStop}%
\bibitem [{\citenamefont {Bargiacchi}\ \emph {et~al.}(2023)\citenamefont
  {Bargiacchi}, \citenamefont {Dainotti},\ and\ \citenamefont
  {Capozziello}}]{Bargiacchi:2023rfd}%
  \BibitemOpen
  \bibfield  {author} {\bibinfo {author} {\bibfnamefont {G.}~\bibnamefont
  {Bargiacchi}}, \bibinfo {author} {\bibfnamefont {M.~G.}\ \bibnamefont
  {Dainotti}}, \ and\ \bibinfo {author} {\bibfnamefont {S.}~\bibnamefont
  {Capozziello}},\ }\href {\doibase 10.1093/mnras/stad2326} {\  (\bibinfo
  {year} {2023}),\ 10.1093/mnras/stad2326},\ \Eprint
  {http://arxiv.org/abs/2307.15359} {arXiv:2307.15359 [astro-ph.CO]}
  \BibitemShut {NoStop}%
\bibitem [{\citenamefont {Rezaei}\ \emph {et~al.}(2020)\citenamefont {Rezaei},
  \citenamefont {Pour-Ojaghi},\ and\ \citenamefont
  {Malekjani}}]{Rezaei:2020lfy}%
  \BibitemOpen
  \bibfield  {author} {\bibinfo {author} {\bibfnamefont {M.}~\bibnamefont
  {Rezaei}}, \bibinfo {author} {\bibfnamefont {S.}~\bibnamefont {Pour-Ojaghi}},
  \ and\ \bibinfo {author} {\bibfnamefont {M.}~\bibnamefont {Malekjani}},\
  }\href {\doibase 10.3847/1538-4357/aba517} {\bibfield  {journal} {\bibinfo
  {journal} {Astrophys. J.}\ }\textbf {\bibinfo {volume} {900}},\ \bibinfo
  {pages} {70} (\bibinfo {year} {2020})},\ \Eprint
  {http://arxiv.org/abs/2008.03092} {arXiv:2008.03092 [astro-ph.CO]}
  \BibitemShut {NoStop}%
\bibitem [{\citenamefont {{Foreman-Mackey}}\ \emph {et~al.}(2013)\citenamefont
  {{Foreman-Mackey}}, \citenamefont {{Hogg}}, \citenamefont {{Lang}},\ and\
  \citenamefont {{Goodman}}}]{Foreman:2013}%
  \BibitemOpen
  \bibfield  {author} {\bibinfo {author} {\bibfnamefont {D.}~\bibnamefont
  {{Foreman-Mackey}}}, \bibinfo {author} {\bibfnamefont {D.~W.}\ \bibnamefont
  {{Hogg}}}, \bibinfo {author} {\bibfnamefont {D.}~\bibnamefont {{Lang}}}, \
  and\ \bibinfo {author} {\bibfnamefont {J.}~\bibnamefont {{Goodman}}},\ }\href
  {\doibase 10.1086/670067} {\bibfield  {journal} {\bibinfo  {journal} {PASP}\
  }\textbf {\bibinfo {volume} {125}},\ \bibinfo {pages} {306} (\bibinfo {year}
  {2013})},\ \Eprint {http://arxiv.org/abs/1202.3665} {arXiv:1202.3665
  [astro-ph.IM]} \BibitemShut {NoStop}%
\bibitem [{\citenamefont {Riess}\ \emph {et~al.}(2019)\citenamefont {Riess},
  \citenamefont {Casertano}, \citenamefont {Yuan}, \citenamefont {Macri},\ and\
  \citenamefont {Scolnic}}]{Riess:2019cxk}%
  \BibitemOpen
  \bibfield  {author} {\bibinfo {author} {\bibfnamefont {A.~G.}\ \bibnamefont
  {Riess}}, \bibinfo {author} {\bibfnamefont {S.}~\bibnamefont {Casertano}},
  \bibinfo {author} {\bibfnamefont {W.}~\bibnamefont {Yuan}}, \bibinfo {author}
  {\bibfnamefont {L.~M.}\ \bibnamefont {Macri}}, \ and\ \bibinfo {author}
  {\bibfnamefont {D.}~\bibnamefont {Scolnic}},\ }\href {\doibase
  10.3847/1538-4357/ab1422} {\bibfield  {journal} {\bibinfo  {journal}
  {Astrophys. J.}\ }\textbf {\bibinfo {volume} {876}},\ \bibinfo {pages} {85}
  (\bibinfo {year} {2019})},\ \Eprint {http://arxiv.org/abs/1903.07603}
  {arXiv:1903.07603 [astro-ph.CO]} \BibitemShut {NoStop}%
%%CITATION = ARXIV:1903.07603;%%
\bibitem [{\citenamefont {Moresco}\ \emph {et~al.}(2016)\citenamefont
  {Moresco}, \citenamefont {Pozzetti}, \citenamefont {Cimatti}, \citenamefont
  {Jimenez}, \citenamefont {Maraston}, \citenamefont {Verde}, \citenamefont
  {Thomas}, \citenamefont {Citro}, \citenamefont {Tojeiro},\ and\ \citenamefont
  {Wilkinson}}]{Moresco:2016mzx}%
  \BibitemOpen
  \bibfield  {author} {\bibinfo {author} {\bibfnamefont {M.}~\bibnamefont
  {Moresco}}, \bibinfo {author} {\bibfnamefont {L.}~\bibnamefont {Pozzetti}},
  \bibinfo {author} {\bibfnamefont {A.}~\bibnamefont {Cimatti}}, \bibinfo
  {author} {\bibfnamefont {R.}~\bibnamefont {Jimenez}}, \bibinfo {author}
  {\bibfnamefont {C.}~\bibnamefont {Maraston}}, \bibinfo {author}
  {\bibfnamefont {L.}~\bibnamefont {Verde}}, \bibinfo {author} {\bibfnamefont
  {D.}~\bibnamefont {Thomas}}, \bibinfo {author} {\bibfnamefont
  {A.}~\bibnamefont {Citro}}, \bibinfo {author} {\bibfnamefont
  {R.}~\bibnamefont {Tojeiro}}, \ and\ \bibinfo {author} {\bibfnamefont
  {D.}~\bibnamefont {Wilkinson}},\ }\href {\doibase
  10.1088/1475-7516/2016/05/014} {\bibfield  {journal} {\bibinfo  {journal}
  {JCAP}\ }\textbf {\bibinfo {volume} {1605}},\ \bibinfo {pages} {014}
  (\bibinfo {year} {2016})},\ \Eprint {http://arxiv.org/abs/1601.01701}
  {arXiv:1601.01701 [astro-ph.CO]} \BibitemShut {NoStop}%
%%CITATION = ARXIV:1601.01701;%%
\bibitem [{\citenamefont {Maga\~na}\ \emph {et~al.}(2018)\citenamefont
  {Maga\~na}, \citenamefont {Amante}, \citenamefont {Garcia-Aspeitia},\ and\
  \citenamefont {Motta}}]{Magana:2017}%
  \BibitemOpen
  \bibfield  {author} {\bibinfo {author} {\bibfnamefont {J.}~\bibnamefont
  {Maga\~na}}, \bibinfo {author} {\bibfnamefont {M.~H.}\ \bibnamefont
  {Amante}}, \bibinfo {author} {\bibfnamefont {M.~A.}\ \bibnamefont
  {Garcia-Aspeitia}}, \ and\ \bibinfo {author} {\bibfnamefont {V.}~\bibnamefont
  {Motta}},\ }\href {\doibase 10.1093/mnras/sty260} {\bibfield  {journal}
  {\bibinfo  {journal} {Mon. Not. Roy. Astron. Soc.}\ }\textbf {\bibinfo
  {volume} {476}},\ \bibinfo {pages} {1036} (\bibinfo {year} {2018})},\ \Eprint
  {http://arxiv.org/abs/1706.09848} {arXiv:1706.09848 [astro-ph.CO]}
  \BibitemShut {NoStop}%
%%CITATION = ARXIV:1706.09848;%%
\bibitem [{\citenamefont {Moresco}(2015)}]{Moresco:2015cya}%
  \BibitemOpen
  \bibfield  {author} {\bibinfo {author} {\bibfnamefont {M.}~\bibnamefont
  {Moresco}},\ }\href {\doibase 10.1093/mnrasl/slv037} {\bibfield  {journal}
  {\bibinfo  {journal} {Mon. Not. Roy. Astron. Soc.}\ }\textbf {\bibinfo
  {volume} {450}},\ \bibinfo {pages} {L16} (\bibinfo {year} {2015})},\ \Eprint
  {http://arxiv.org/abs/1503.01116} {arXiv:1503.01116 [astro-ph.CO]}
  \BibitemShut {NoStop}%
\bibitem [{\citenamefont {Scolnic}\ \emph {et~al.}(2018)\citenamefont
  {Scolnic}, \citenamefont {Jones}, \citenamefont {Rest}, \citenamefont {Pan},
  \citenamefont {Chornock}, \citenamefont {Foley}, \citenamefont {Huber},
  \citenamefont {Kessler}, \citenamefont {Narayan}, \citenamefont {Riess},
  \citenamefont {Rodney}, \citenamefont {Berger}, \citenamefont {Brout},
  \citenamefont {Challis}, \citenamefont {Drout}, \citenamefont {Finkbeiner},
  \citenamefont {Lunnan}, \citenamefont {Kirshner}, \citenamefont {Sanders},
  \citenamefont {Schlafly}, \citenamefont {Smartt}, \citenamefont {Stubbs},
  \citenamefont {Tonry}, \citenamefont {Wood-Vasey}, \citenamefont {Foley},
  \citenamefont {Hand}, \citenamefont {Johnson}, \citenamefont {Burgett},
  \citenamefont {Chambers}, \citenamefont {Draper}, \citenamefont {Hodapp},
  \citenamefont {Kaiser}, \citenamefont {Kudritzki}, \citenamefont {Magnier},
  \citenamefont {Metcalfe}, \citenamefont {Bresolin}, \citenamefont {Gall},
  \citenamefont {Kotak}, \citenamefont {McCrum},\ and\ \citenamefont
  {Smith}}]{Scolnic2018-qf}%
  \BibitemOpen
  \bibfield  {author} {\bibinfo {author} {\bibfnamefont {D.~M.}\ \bibnamefont
  {Scolnic}}, \bibinfo {author} {\bibfnamefont {D.~O.}\ \bibnamefont {Jones}},
  \bibinfo {author} {\bibfnamefont {A.}~\bibnamefont {Rest}}, \bibinfo {author}
  {\bibfnamefont {Y.~C.}\ \bibnamefont {Pan}}, \bibinfo {author} {\bibfnamefont
  {R.}~\bibnamefont {Chornock}}, \bibinfo {author} {\bibfnamefont {R.~J.}\
  \bibnamefont {Foley}}, \bibinfo {author} {\bibfnamefont {M.~E.}\ \bibnamefont
  {Huber}}, \bibinfo {author} {\bibfnamefont {R.}~\bibnamefont {Kessler}},
  \bibinfo {author} {\bibfnamefont {G.}~\bibnamefont {Narayan}}, \bibinfo
  {author} {\bibfnamefont {A.~G.}\ \bibnamefont {Riess}}, \bibinfo {author}
  {\bibfnamefont {S.}~\bibnamefont {Rodney}}, \bibinfo {author} {\bibfnamefont
  {E.}~\bibnamefont {Berger}}, \bibinfo {author} {\bibfnamefont {D.~J.}\
  \bibnamefont {Brout}}, \bibinfo {author} {\bibfnamefont {P.~J.}\ \bibnamefont
  {Challis}}, \bibinfo {author} {\bibfnamefont {M.}~\bibnamefont {Drout}},
  \bibinfo {author} {\bibfnamefont {D.}~\bibnamefont {Finkbeiner}}, \bibinfo
  {author} {\bibfnamefont {R.}~\bibnamefont {Lunnan}}, \bibinfo {author}
  {\bibfnamefont {R.~P.}\ \bibnamefont {Kirshner}}, \bibinfo {author}
  {\bibfnamefont {N.~E.}\ \bibnamefont {Sanders}}, \bibinfo {author}
  {\bibfnamefont {E.}~\bibnamefont {Schlafly}}, \bibinfo {author}
  {\bibfnamefont {S.}~\bibnamefont {Smartt}}, \bibinfo {author} {\bibfnamefont
  {C.~W.}\ \bibnamefont {Stubbs}}, \bibinfo {author} {\bibfnamefont
  {J.}~\bibnamefont {Tonry}}, \bibinfo {author} {\bibfnamefont {W.~M.}\
  \bibnamefont {Wood-Vasey}}, \bibinfo {author} {\bibfnamefont
  {M.}~\bibnamefont {Foley}}, \bibinfo {author} {\bibfnamefont
  {J.}~\bibnamefont {Hand}}, \bibinfo {author} {\bibfnamefont {E.}~\bibnamefont
  {Johnson}}, \bibinfo {author} {\bibfnamefont {W.~S.}\ \bibnamefont
  {Burgett}}, \bibinfo {author} {\bibfnamefont {K.~C.}\ \bibnamefont
  {Chambers}}, \bibinfo {author} {\bibfnamefont {P.~W.}\ \bibnamefont
  {Draper}}, \bibinfo {author} {\bibfnamefont {K.~W.}\ \bibnamefont {Hodapp}},
  \bibinfo {author} {\bibfnamefont {N.}~\bibnamefont {Kaiser}}, \bibinfo
  {author} {\bibfnamefont {R.~P.}\ \bibnamefont {Kudritzki}}, \bibinfo {author}
  {\bibfnamefont {E.~A.}\ \bibnamefont {Magnier}}, \bibinfo {author}
  {\bibfnamefont {N.}~\bibnamefont {Metcalfe}}, \bibinfo {author}
  {\bibfnamefont {F.}~\bibnamefont {Bresolin}}, \bibinfo {author}
  {\bibfnamefont {E.}~\bibnamefont {Gall}}, \bibinfo {author} {\bibfnamefont
  {R.}~\bibnamefont {Kotak}}, \bibinfo {author} {\bibfnamefont
  {M.}~\bibnamefont {McCrum}}, \ and\ \bibinfo {author} {\bibfnamefont {K.~W.}\
  \bibnamefont {Smith}},\ }\href@noop {} {\bibfield  {journal} {\bibinfo
  {journal} {Astrophys. J.}\ }\textbf {\bibinfo {volume} {859}},\ \bibinfo
  {pages} {101} (\bibinfo {year} {2018})}\BibitemShut {NoStop}%
\bibitem [{\citenamefont {Brout}\ \emph {et~al.}(2022)\citenamefont {Brout},
  \citenamefont {Scolnic}, \citenamefont {Popovic}, \citenamefont {Riess},
  \citenamefont {Carr}, \citenamefont {Zuntz}, \citenamefont {Kessler},
  \citenamefont {Davis}, \citenamefont {Hinton}, \citenamefont {Jones},
  \citenamefont {Kenworthy}, \citenamefont {Peterson}, \citenamefont {Said},
  \citenamefont {Taylor}, \citenamefont {Ali}, \citenamefont {Armstrong},
  \citenamefont {Charvu}, \citenamefont {Dwomoh}, \citenamefont {Meldorf},
  \citenamefont {Palmese}, \citenamefont {Qu}, \citenamefont {Rose},
  \citenamefont {Sanchez}, \citenamefont {Stubbs}, \citenamefont {Vincenzi},
  \citenamefont {Wood}, \citenamefont {Brown}, \citenamefont {Chen},
  \citenamefont {Chambers}, \citenamefont {Coulter}, \citenamefont {Dai},
  \citenamefont {Dimitriadis}, \citenamefont {Filippenko}, \citenamefont
  {Foley}, \citenamefont {Jha}, \citenamefont {Kelsey}, \citenamefont
  {Kirshner}, \citenamefont {Möller}, \citenamefont {Muir}, \citenamefont
  {Nadathur}, \citenamefont {Pan}, \citenamefont {Rest}, \citenamefont
  {Rojas-Bravo}, \citenamefont {Sako}, \citenamefont {Siebert}, \citenamefont
  {Smith}, \citenamefont {Stahl},\ and\ \citenamefont {Wiseman}}]{Brout_2022}%
  \BibitemOpen
  \bibfield  {author} {\bibinfo {author} {\bibfnamefont {D.}~\bibnamefont
  {Brout}}, \bibinfo {author} {\bibfnamefont {D.}~\bibnamefont {Scolnic}},
  \bibinfo {author} {\bibfnamefont {B.}~\bibnamefont {Popovic}}, \bibinfo
  {author} {\bibfnamefont {A.~G.}\ \bibnamefont {Riess}}, \bibinfo {author}
  {\bibfnamefont {A.}~\bibnamefont {Carr}}, \bibinfo {author} {\bibfnamefont
  {J.}~\bibnamefont {Zuntz}}, \bibinfo {author} {\bibfnamefont
  {R.}~\bibnamefont {Kessler}}, \bibinfo {author} {\bibfnamefont {T.~M.}\
  \bibnamefont {Davis}}, \bibinfo {author} {\bibfnamefont {S.}~\bibnamefont
  {Hinton}}, \bibinfo {author} {\bibfnamefont {D.}~\bibnamefont {Jones}},
  \bibinfo {author} {\bibfnamefont {W.~D.}\ \bibnamefont {Kenworthy}}, \bibinfo
  {author} {\bibfnamefont {E.~R.}\ \bibnamefont {Peterson}}, \bibinfo {author}
  {\bibfnamefont {K.}~\bibnamefont {Said}}, \bibinfo {author} {\bibfnamefont
  {G.}~\bibnamefont {Taylor}}, \bibinfo {author} {\bibfnamefont
  {N.}~\bibnamefont {Ali}}, \bibinfo {author} {\bibfnamefont {P.}~\bibnamefont
  {Armstrong}}, \bibinfo {author} {\bibfnamefont {P.}~\bibnamefont {Charvu}},
  \bibinfo {author} {\bibfnamefont {A.}~\bibnamefont {Dwomoh}}, \bibinfo
  {author} {\bibfnamefont {C.}~\bibnamefont {Meldorf}}, \bibinfo {author}
  {\bibfnamefont {A.}~\bibnamefont {Palmese}}, \bibinfo {author} {\bibfnamefont
  {H.}~\bibnamefont {Qu}}, \bibinfo {author} {\bibfnamefont {B.~M.}\
  \bibnamefont {Rose}}, \bibinfo {author} {\bibfnamefont {B.}~\bibnamefont
  {Sanchez}}, \bibinfo {author} {\bibfnamefont {C.~W.}\ \bibnamefont {Stubbs}},
  \bibinfo {author} {\bibfnamefont {M.}~\bibnamefont {Vincenzi}}, \bibinfo
  {author} {\bibfnamefont {C.~M.}\ \bibnamefont {Wood}}, \bibinfo {author}
  {\bibfnamefont {P.~J.}\ \bibnamefont {Brown}}, \bibinfo {author}
  {\bibfnamefont {R.}~\bibnamefont {Chen}}, \bibinfo {author} {\bibfnamefont
  {K.}~\bibnamefont {Chambers}}, \bibinfo {author} {\bibfnamefont {D.~A.}\
  \bibnamefont {Coulter}}, \bibinfo {author} {\bibfnamefont {M.}~\bibnamefont
  {Dai}}, \bibinfo {author} {\bibfnamefont {G.}~\bibnamefont {Dimitriadis}},
  \bibinfo {author} {\bibfnamefont {A.~V.}\ \bibnamefont {Filippenko}},
  \bibinfo {author} {\bibfnamefont {R.~J.}\ \bibnamefont {Foley}}, \bibinfo
  {author} {\bibfnamefont {S.~W.}\ \bibnamefont {Jha}}, \bibinfo {author}
  {\bibfnamefont {L.}~\bibnamefont {Kelsey}}, \bibinfo {author} {\bibfnamefont
  {R.~P.}\ \bibnamefont {Kirshner}}, \bibinfo {author} {\bibfnamefont
  {A.}~\bibnamefont {Möller}}, \bibinfo {author} {\bibfnamefont
  {J.}~\bibnamefont {Muir}}, \bibinfo {author} {\bibfnamefont {S.}~\bibnamefont
  {Nadathur}}, \bibinfo {author} {\bibfnamefont {Y.-C.}\ \bibnamefont {Pan}},
  \bibinfo {author} {\bibfnamefont {A.}~\bibnamefont {Rest}}, \bibinfo {author}
  {\bibfnamefont {C.}~\bibnamefont {Rojas-Bravo}}, \bibinfo {author}
  {\bibfnamefont {M.}~\bibnamefont {Sako}}, \bibinfo {author} {\bibfnamefont
  {M.~R.}\ \bibnamefont {Siebert}}, \bibinfo {author} {\bibfnamefont
  {M.}~\bibnamefont {Smith}}, \bibinfo {author} {\bibfnamefont {B.~E.}\
  \bibnamefont {Stahl}}, \ and\ \bibinfo {author} {\bibfnamefont
  {P.}~\bibnamefont {Wiseman}},\ }\href {\doibase 10.3847/1538-4357/ac8e04}
  {\bibfield  {journal} {\bibinfo  {journal} {The Astrophysical Journal}\
  }\textbf {\bibinfo {volume} {938}},\ \bibinfo {pages} {110} (\bibinfo {year}
  {2022})}\BibitemShut {NoStop}%
\bibitem [{\citenamefont {Conley}\ \emph {et~al.}(2010)\citenamefont {Conley},
  \citenamefont {Guy}, \citenamefont {Sullivan}, \citenamefont {Regnault},
  \citenamefont {Astier}, \citenamefont {Balland}, \citenamefont {Basa},
  \citenamefont {Carlberg}, \citenamefont {Fouchez}, \citenamefont {Hardin},
  \citenamefont {Hook}, \citenamefont {Howell}, \citenamefont {Pain},
  \citenamefont {Palanque-Delabrouille}, \citenamefont {Perrett}, \citenamefont
  {Pritchet}, \citenamefont {Rich}, \citenamefont {Ruhlmann-Kleider},
  \citenamefont {Balam}, \citenamefont {Baumont}, \citenamefont {Ellis},
  \citenamefont {Fabbro}, \citenamefont {Fakhouri}, \citenamefont {Fourmanoit},
  \citenamefont {González-Gaitán}, \citenamefont {Graham}, \citenamefont
  {Hudson}, \citenamefont {Hsiao}, \citenamefont {Kronborg}, \citenamefont
  {Lidman}, \citenamefont {Mourao}, \citenamefont {Neill}, \citenamefont
  {Perlmutter}, \citenamefont {Ripoche}, \citenamefont {Suzuki},\ and\
  \citenamefont {Walker}}]{Conley_2010}%
  \BibitemOpen
  \bibfield  {author} {\bibinfo {author} {\bibfnamefont {A.}~\bibnamefont
  {Conley}}, \bibinfo {author} {\bibfnamefont {J.}~\bibnamefont {Guy}},
  \bibinfo {author} {\bibfnamefont {M.}~\bibnamefont {Sullivan}}, \bibinfo
  {author} {\bibfnamefont {N.}~\bibnamefont {Regnault}}, \bibinfo {author}
  {\bibfnamefont {P.}~\bibnamefont {Astier}}, \bibinfo {author} {\bibfnamefont
  {C.}~\bibnamefont {Balland}}, \bibinfo {author} {\bibfnamefont
  {S.}~\bibnamefont {Basa}}, \bibinfo {author} {\bibfnamefont {R.~G.}\
  \bibnamefont {Carlberg}}, \bibinfo {author} {\bibfnamefont {D.}~\bibnamefont
  {Fouchez}}, \bibinfo {author} {\bibfnamefont {D.}~\bibnamefont {Hardin}},
  \bibinfo {author} {\bibfnamefont {I.~M.}\ \bibnamefont {Hook}}, \bibinfo
  {author} {\bibfnamefont {D.~A.}\ \bibnamefont {Howell}}, \bibinfo {author}
  {\bibfnamefont {R.}~\bibnamefont {Pain}}, \bibinfo {author} {\bibfnamefont
  {N.}~\bibnamefont {Palanque-Delabrouille}}, \bibinfo {author} {\bibfnamefont
  {K.~M.}\ \bibnamefont {Perrett}}, \bibinfo {author} {\bibfnamefont {C.~J.}\
  \bibnamefont {Pritchet}}, \bibinfo {author} {\bibfnamefont {J.}~\bibnamefont
  {Rich}}, \bibinfo {author} {\bibfnamefont {V.}~\bibnamefont
  {Ruhlmann-Kleider}}, \bibinfo {author} {\bibfnamefont {D.}~\bibnamefont
  {Balam}}, \bibinfo {author} {\bibfnamefont {S.}~\bibnamefont {Baumont}},
  \bibinfo {author} {\bibfnamefont {R.~S.}\ \bibnamefont {Ellis}}, \bibinfo
  {author} {\bibfnamefont {S.}~\bibnamefont {Fabbro}}, \bibinfo {author}
  {\bibfnamefont {H.~K.}\ \bibnamefont {Fakhouri}}, \bibinfo {author}
  {\bibfnamefont {N.}~\bibnamefont {Fourmanoit}}, \bibinfo {author}
  {\bibfnamefont {S.}~\bibnamefont {González-Gaitán}}, \bibinfo {author}
  {\bibfnamefont {M.~L.}\ \bibnamefont {Graham}}, \bibinfo {author}
  {\bibfnamefont {M.~J.}\ \bibnamefont {Hudson}}, \bibinfo {author}
  {\bibfnamefont {E.}~\bibnamefont {Hsiao}}, \bibinfo {author} {\bibfnamefont
  {T.}~\bibnamefont {Kronborg}}, \bibinfo {author} {\bibfnamefont
  {C.}~\bibnamefont {Lidman}}, \bibinfo {author} {\bibfnamefont {A.~M.}\
  \bibnamefont {Mourao}}, \bibinfo {author} {\bibfnamefont {J.~D.}\
  \bibnamefont {Neill}}, \bibinfo {author} {\bibfnamefont {S.}~\bibnamefont
  {Perlmutter}}, \bibinfo {author} {\bibfnamefont {P.}~\bibnamefont {Ripoche}},
  \bibinfo {author} {\bibfnamefont {N.}~\bibnamefont {Suzuki}}, \ and\ \bibinfo
  {author} {\bibfnamefont {E.~S.}\ \bibnamefont {Walker}},\ }\href {\doibase
  10.1088/0067-0049/192/1/1} {\bibfield  {journal} {\bibinfo  {journal} {The
  Astrophysical Journal Supplement Series}\ }\textbf {\bibinfo {volume}
  {192}},\ \bibinfo {pages} {1} (\bibinfo {year} {2010})}\BibitemShut {NoStop}%
\bibitem [{\citenamefont {Alam}\ \emph {et~al.}(2021)\citenamefont {Alam},
  \citenamefont {Aubert}, \citenamefont {Avila}, \citenamefont {Balland},
  \citenamefont {Bautista}, \citenamefont {Bershady}, \citenamefont {Bizyaev},
  \citenamefont {Blanton}, \citenamefont {Bolton}, \citenamefont {Bovy},
  \citenamefont {Brinkmann}, \citenamefont {Brownstein}, \citenamefont
  {Burtin}, \citenamefont {Chabanier}, \citenamefont {Chapman}, \citenamefont
  {Choi}, \citenamefont {Chuang}, \citenamefont {Comparat}, \citenamefont
  {Cousinou}, \citenamefont {Cuceu}, \citenamefont {Dawson}, \citenamefont
  {de~la Torre}, \citenamefont {de~Mattia}, \citenamefont {de~Sainte~Agathe},
  \citenamefont {du~Mas~des Bourboux}, \citenamefont {Escoffier}, \citenamefont
  {Etourneau}, \citenamefont {Farr}, \citenamefont {Font-Ribera}, \citenamefont
  {Frinchaboy}, \citenamefont {Fromenteau}, \citenamefont {Gil-Mar{\'{\i}}n},
  \citenamefont {Goff}, \citenamefont {Gonzalez-Morales}, \citenamefont
  {Gonzalez-Perez}, \citenamefont {Grabowski}, \citenamefont {Guy},
  \citenamefont {Hawken}, \citenamefont {Hou}, \citenamefont {Kong},
  \citenamefont {Parker}, \citenamefont {Klaene}, \citenamefont {Kneib},
  \citenamefont {Lin}, \citenamefont {Long}, \citenamefont {Lyke},
  \citenamefont {de~la Macorra}, \citenamefont {Martini}, \citenamefont
  {Masters}, \citenamefont {Mohammad}, \citenamefont {Moon}, \citenamefont
  {Mueller}, \citenamefont {Mu{\~{n}}oz-Guti{\'{e}}rrez}, \citenamefont
  {Myers}, \citenamefont {Nadathur}, \citenamefont {Neveux}, \citenamefont
  {Newman}, \citenamefont {Noterdaeme}, \citenamefont {Oravetz}, \citenamefont
  {Oravetz}, \citenamefont {Palanque-Delabrouille}, \citenamefont {Pan},
  \citenamefont {Paviot}, \citenamefont {Percival}, \citenamefont
  {P{\'{e}}rez-R{\`{a}}fols}, \citenamefont {Petitjean}, \citenamefont {Pieri},
  \citenamefont {Prakash}, \citenamefont {Raichoor}, \citenamefont {Ravoux},
  \citenamefont {Rezaie}, \citenamefont {Rich}, \citenamefont {Ross},
  \citenamefont {Rossi}, \citenamefont {Ruggeri}, \citenamefont
  {Ruhlmann-Kleider}, \citenamefont {S{\'{a}}nchez}, \citenamefont
  {S{\'{a}}nchez}, \citenamefont {S{\'{a}}nchez-Gallego}, \citenamefont
  {Sayres}, \citenamefont {Schneider}, \citenamefont {Seo}, \citenamefont
  {Shafieloo}, \citenamefont {Slosar}, \citenamefont {Smith}, \citenamefont
  {Stermer}, \citenamefont {Tamone}, \citenamefont {Tinker}, \citenamefont
  {Tojeiro}, \citenamefont {Vargas-Maga{\~{n}}a}, \citenamefont {Variu},
  \citenamefont {Wang}, \citenamefont {Weaver}, \citenamefont {Weijmans},
  \citenamefont {Y{\`{e}}che}, \citenamefont {Zarrouk}, \citenamefont {Zhao},
  \citenamefont {Zhao},\ and\ \citenamefont {Zheng}}]{Alam_2021}%
  \BibitemOpen
  \bibfield  {author} {\bibinfo {author} {\bibfnamefont {S.}~\bibnamefont
  {Alam}}, \bibinfo {author} {\bibfnamefont {M.}~\bibnamefont {Aubert}},
  \bibinfo {author} {\bibfnamefont {S.}~\bibnamefont {Avila}}, \bibinfo
  {author} {\bibfnamefont {C.}~\bibnamefont {Balland}}, \bibinfo {author}
  {\bibfnamefont {J.~E.}\ \bibnamefont {Bautista}}, \bibinfo {author}
  {\bibfnamefont {M.~A.}\ \bibnamefont {Bershady}}, \bibinfo {author}
  {\bibfnamefont {D.}~\bibnamefont {Bizyaev}}, \bibinfo {author} {\bibfnamefont
  {M.~R.}\ \bibnamefont {Blanton}}, \bibinfo {author} {\bibfnamefont {A.~S.}\
  \bibnamefont {Bolton}}, \bibinfo {author} {\bibfnamefont {J.}~\bibnamefont
  {Bovy}}, \bibinfo {author} {\bibfnamefont {J.}~\bibnamefont {Brinkmann}},
  \bibinfo {author} {\bibfnamefont {J.~R.}\ \bibnamefont {Brownstein}},
  \bibinfo {author} {\bibfnamefont {E.}~\bibnamefont {Burtin}}, \bibinfo
  {author} {\bibfnamefont {S.}~\bibnamefont {Chabanier}}, \bibinfo {author}
  {\bibfnamefont {M.~J.}\ \bibnamefont {Chapman}}, \bibinfo {author}
  {\bibfnamefont {P.~D.}\ \bibnamefont {Choi}}, \bibinfo {author}
  {\bibfnamefont {C.-H.}\ \bibnamefont {Chuang}}, \bibinfo {author}
  {\bibfnamefont {J.}~\bibnamefont {Comparat}}, \bibinfo {author}
  {\bibfnamefont {M.-C.}\ \bibnamefont {Cousinou}}, \bibinfo {author}
  {\bibfnamefont {A.}~\bibnamefont {Cuceu}}, \bibinfo {author} {\bibfnamefont
  {K.~S.}\ \bibnamefont {Dawson}}, \bibinfo {author} {\bibfnamefont
  {S.}~\bibnamefont {de~la Torre}}, \bibinfo {author} {\bibfnamefont
  {A.}~\bibnamefont {de~Mattia}}, \bibinfo {author} {\bibfnamefont
  {V.}~\bibnamefont {de~Sainte~Agathe}}, \bibinfo {author} {\bibfnamefont
  {H.}~\bibnamefont {du~Mas~des Bourboux}}, \bibinfo {author} {\bibfnamefont
  {S.}~\bibnamefont {Escoffier}}, \bibinfo {author} {\bibfnamefont
  {T.}~\bibnamefont {Etourneau}}, \bibinfo {author} {\bibfnamefont
  {J.}~\bibnamefont {Farr}}, \bibinfo {author} {\bibfnamefont {A.}~\bibnamefont
  {Font-Ribera}}, \bibinfo {author} {\bibfnamefont {P.~M.}\ \bibnamefont
  {Frinchaboy}}, \bibinfo {author} {\bibfnamefont {S.}~\bibnamefont
  {Fromenteau}}, \bibinfo {author} {\bibfnamefont {H.}~\bibnamefont
  {Gil-Mar{\'{\i}}n}}, \bibinfo {author} {\bibfnamefont {J.-M.~L.}\
  \bibnamefont {Goff}}, \bibinfo {author} {\bibfnamefont {A.~X.}\ \bibnamefont
  {Gonzalez-Morales}}, \bibinfo {author} {\bibfnamefont {V.}~\bibnamefont
  {Gonzalez-Perez}}, \bibinfo {author} {\bibfnamefont {K.}~\bibnamefont
  {Grabowski}}, \bibinfo {author} {\bibfnamefont {J.}~\bibnamefont {Guy}},
  \bibinfo {author} {\bibfnamefont {A.~J.}\ \bibnamefont {Hawken}}, \bibinfo
  {author} {\bibfnamefont {J.}~\bibnamefont {Hou}}, \bibinfo {author}
  {\bibfnamefont {H.}~\bibnamefont {Kong}}, \bibinfo {author} {\bibfnamefont
  {J.}~\bibnamefont {Parker}}, \bibinfo {author} {\bibfnamefont
  {M.}~\bibnamefont {Klaene}}, \bibinfo {author} {\bibfnamefont {J.-P.}\
  \bibnamefont {Kneib}}, \bibinfo {author} {\bibfnamefont {S.}~\bibnamefont
  {Lin}}, \bibinfo {author} {\bibfnamefont {D.}~\bibnamefont {Long}}, \bibinfo
  {author} {\bibfnamefont {B.~W.}\ \bibnamefont {Lyke}}, \bibinfo {author}
  {\bibfnamefont {A.}~\bibnamefont {de~la Macorra}}, \bibinfo {author}
  {\bibfnamefont {P.}~\bibnamefont {Martini}}, \bibinfo {author} {\bibfnamefont
  {K.}~\bibnamefont {Masters}}, \bibinfo {author} {\bibfnamefont {F.~G.}\
  \bibnamefont {Mohammad}}, \bibinfo {author} {\bibfnamefont {J.}~\bibnamefont
  {Moon}}, \bibinfo {author} {\bibfnamefont {E.-M.}\ \bibnamefont {Mueller}},
  \bibinfo {author} {\bibfnamefont {A.}~\bibnamefont
  {Mu{\~{n}}oz-Guti{\'{e}}rrez}}, \bibinfo {author} {\bibfnamefont {A.~D.}\
  \bibnamefont {Myers}}, \bibinfo {author} {\bibfnamefont {S.}~\bibnamefont
  {Nadathur}}, \bibinfo {author} {\bibfnamefont {R.}~\bibnamefont {Neveux}},
  \bibinfo {author} {\bibfnamefont {J.~A.}\ \bibnamefont {Newman}}, \bibinfo
  {author} {\bibfnamefont {P.}~\bibnamefont {Noterdaeme}}, \bibinfo {author}
  {\bibfnamefont {A.}~\bibnamefont {Oravetz}}, \bibinfo {author} {\bibfnamefont
  {D.}~\bibnamefont {Oravetz}}, \bibinfo {author} {\bibfnamefont
  {N.}~\bibnamefont {Palanque-Delabrouille}}, \bibinfo {author} {\bibfnamefont
  {K.}~\bibnamefont {Pan}}, \bibinfo {author} {\bibfnamefont {R.}~\bibnamefont
  {Paviot}}, \bibinfo {author} {\bibfnamefont {W.~J.}\ \bibnamefont
  {Percival}}, \bibinfo {author} {\bibfnamefont {I.}~\bibnamefont
  {P{\'{e}}rez-R{\`{a}}fols}}, \bibinfo {author} {\bibfnamefont
  {P.}~\bibnamefont {Petitjean}}, \bibinfo {author} {\bibfnamefont {M.~M.}\
  \bibnamefont {Pieri}}, \bibinfo {author} {\bibfnamefont {A.}~\bibnamefont
  {Prakash}}, \bibinfo {author} {\bibfnamefont {A.}~\bibnamefont {Raichoor}},
  \bibinfo {author} {\bibfnamefont {C.}~\bibnamefont {Ravoux}}, \bibinfo
  {author} {\bibfnamefont {M.}~\bibnamefont {Rezaie}}, \bibinfo {author}
  {\bibfnamefont {J.}~\bibnamefont {Rich}}, \bibinfo {author} {\bibfnamefont
  {A.~J.}\ \bibnamefont {Ross}}, \bibinfo {author} {\bibfnamefont
  {G.}~\bibnamefont {Rossi}}, \bibinfo {author} {\bibfnamefont
  {R.}~\bibnamefont {Ruggeri}}, \bibinfo {author} {\bibfnamefont
  {V.}~\bibnamefont {Ruhlmann-Kleider}}, \bibinfo {author} {\bibfnamefont
  {A.~G.}\ \bibnamefont {S{\'{a}}nchez}}, \bibinfo {author} {\bibfnamefont
  {F.~J.}\ \bibnamefont {S{\'{a}}nchez}}, \bibinfo {author} {\bibfnamefont
  {J.~R.}\ \bibnamefont {S{\'{a}}nchez-Gallego}}, \bibinfo {author}
  {\bibfnamefont {C.}~\bibnamefont {Sayres}}, \bibinfo {author} {\bibfnamefont
  {D.~P.}\ \bibnamefont {Schneider}}, \bibinfo {author} {\bibfnamefont {H.-J.}\
  \bibnamefont {Seo}}, \bibinfo {author} {\bibfnamefont {A.}~\bibnamefont
  {Shafieloo}}, \bibinfo {author} {\bibfnamefont {A.}~\bibnamefont {Slosar}},
  \bibinfo {author} {\bibfnamefont {A.}~\bibnamefont {Smith}}, \bibinfo
  {author} {\bibfnamefont {J.}~\bibnamefont {Stermer}}, \bibinfo {author}
  {\bibfnamefont {A.}~\bibnamefont {Tamone}}, \bibinfo {author} {\bibfnamefont
  {J.~L.}\ \bibnamefont {Tinker}}, \bibinfo {author} {\bibfnamefont
  {R.}~\bibnamefont {Tojeiro}}, \bibinfo {author} {\bibfnamefont
  {M.}~\bibnamefont {Vargas-Maga{\~{n}}a}}, \bibinfo {author} {\bibfnamefont
  {A.}~\bibnamefont {Variu}}, \bibinfo {author} {\bibfnamefont
  {Y.}~\bibnamefont {Wang}}, \bibinfo {author} {\bibfnamefont {B.~A.}\
  \bibnamefont {Weaver}}, \bibinfo {author} {\bibfnamefont {A.-M.}\
  \bibnamefont {Weijmans}}, \bibinfo {author} {\bibfnamefont {C.}~\bibnamefont
  {Y{\`{e}}che}}, \bibinfo {author} {\bibfnamefont {P.}~\bibnamefont
  {Zarrouk}}, \bibinfo {author} {\bibfnamefont {C.}~\bibnamefont {Zhao}},
  \bibinfo {author} {\bibfnamefont {G.-B.}\ \bibnamefont {Zhao}}, \ and\
  \bibinfo {author} {\bibfnamefont {Z.}~\bibnamefont {Zheng}},\ }\href
  {\doibase 10.1103/physrevd.103.083533} {\bibfield  {journal} {\bibinfo
  {journal} {Physical Review D}\ }\textbf {\bibinfo {volume} {103}} (\bibinfo
  {year} {2021}),\ 10.1103/physrevd.103.083533}\BibitemShut {NoStop}%
\bibitem [{\citenamefont {Eisenstein}\ \emph {et~al.}(2005)\citenamefont
  {Eisenstein} \emph {et~al.}}]{Wigglez:Eisenstein2005}%
  \BibitemOpen
  \bibfield  {author} {\bibinfo {author} {\bibfnamefont {D.~J.}\ \bibnamefont
  {Eisenstein}} \emph {et~al.} (\bibinfo {collaboration} {SDSS}),\ }\href
  {\doibase 10.1086/466512} {\bibfield  {journal} {\bibinfo  {journal}
  {Astrophys. J.}\ }\textbf {\bibinfo {volume} {633}},\ \bibinfo {pages} {560}
  (\bibinfo {year} {2005})},\ \Eprint {http://arxiv.org/abs/astro-ph/0501171}
  {arXiv:astro-ph/0501171 [astro-ph]} \BibitemShut {NoStop}%
%%CITATION = ASTRO-PH/0501171;%%
\bibitem [{\citenamefont {Padmanabhan}\ and\ \citenamefont
  {Loeb}(2023)}]{Padmanabhan:2023esp}%
  \BibitemOpen
  \bibfield  {author} {\bibinfo {author} {\bibfnamefont {H.}~\bibnamefont
  {Padmanabhan}}\ and\ \bibinfo {author} {\bibfnamefont {A.}~\bibnamefont
  {Loeb}},\ }\href {\doibase 10.3847/2041-8213/acea7a} {\bibfield  {journal}
  {\bibinfo  {journal} {Astrophys. J. Lett.}\ }\textbf {\bibinfo {volume}
  {953}},\ \bibinfo {pages} {L4} (\bibinfo {year} {2023})},\ \Eprint
  {http://arxiv.org/abs/2306.04684} {arXiv:2306.04684 [astro-ph.CO]}
  \BibitemShut {NoStop}%
\bibitem [{\citenamefont {{Gonz{\'a}lez-Mor{\'a}n}}\ \emph
  {et~al.}(2019)\citenamefont {{Gonz{\'a}lez-Mor{\'a}n}}, \citenamefont
  {{Ch{\'a}vez}}, \citenamefont {{Terlevich}}, \citenamefont {{Terlevich}},
  \citenamefont {{Bresolin}}, \citenamefont {{Fern{\'a}ndez-Arenas}},
  \citenamefont {{Plionis}}, \citenamefont {{Basilakos}}, \citenamefont
  {{Melnick}},\ and\ \citenamefont {{Telles}}}]{GonzalezMoran2019}%
  \BibitemOpen
  \bibfield  {author} {\bibinfo {author} {\bibfnamefont {A.~L.}\ \bibnamefont
  {{Gonz{\'a}lez-Mor{\'a}n}}}, \bibinfo {author} {\bibfnamefont
  {R.}~\bibnamefont {{Ch{\'a}vez}}}, \bibinfo {author} {\bibfnamefont
  {R.}~\bibnamefont {{Terlevich}}}, \bibinfo {author} {\bibfnamefont
  {E.}~\bibnamefont {{Terlevich}}}, \bibinfo {author} {\bibfnamefont
  {F.}~\bibnamefont {{Bresolin}}}, \bibinfo {author} {\bibfnamefont
  {D.}~\bibnamefont {{Fern{\'a}ndez-Arenas}}}, \bibinfo {author} {\bibfnamefont
  {M.}~\bibnamefont {{Plionis}}}, \bibinfo {author} {\bibfnamefont
  {S.}~\bibnamefont {{Basilakos}}}, \bibinfo {author} {\bibfnamefont
  {J.}~\bibnamefont {{Melnick}}}, \ and\ \bibinfo {author} {\bibfnamefont
  {E.}~\bibnamefont {{Telles}}},\ }\href {\doibase 10.1093/mnras/stz1577}
  {\bibfield  {journal} {\bibinfo  {journal} {Monthly Notices of the Royal
  Astronomical Society}\ }\textbf {\bibinfo {volume} {487}},\ \bibinfo {pages}
  {4669} (\bibinfo {year} {2019})},\ \Eprint {http://arxiv.org/abs/1906.02195}
  {arXiv:1906.02195 [astro-ph.GA]} \BibitemShut {NoStop}%
\bibitem [{\citenamefont {Gonz\'alez-Mor\'an}\ \emph
  {et~al.}(2021)\citenamefont {Gonz\'alez-Mor\'an}, \citenamefont {Ch\'avez},
  \citenamefont {Terlevich}, \citenamefont {Terlevich}, \citenamefont
  {Fern\'andez-Arenas}, \citenamefont {Bresolin}, \citenamefont {Plionis},
  \citenamefont {Melnick}, \citenamefont {Basilakos},\ and\ \citenamefont
  {Telles}}]{Gonzalez-Moran:2021drc}%
  \BibitemOpen
  \bibfield  {author} {\bibinfo {author} {\bibfnamefont {A.~L.}\ \bibnamefont
  {Gonz\'alez-Mor\'an}}, \bibinfo {author} {\bibfnamefont {R.}~\bibnamefont
  {Ch\'avez}}, \bibinfo {author} {\bibfnamefont {E.}~\bibnamefont {Terlevich}},
  \bibinfo {author} {\bibfnamefont {R.}~\bibnamefont {Terlevich}}, \bibinfo
  {author} {\bibfnamefont {D.}~\bibnamefont {Fern\'andez-Arenas}}, \bibinfo
  {author} {\bibfnamefont {F.}~\bibnamefont {Bresolin}}, \bibinfo {author}
  {\bibfnamefont {M.}~\bibnamefont {Plionis}}, \bibinfo {author} {\bibfnamefont
  {J.}~\bibnamefont {Melnick}}, \bibinfo {author} {\bibfnamefont
  {S.}~\bibnamefont {Basilakos}}, \ and\ \bibinfo {author} {\bibfnamefont
  {E.}~\bibnamefont {Telles}},\ }\href {\doibase 10.1093/mnras/stab1385}
  {\bibfield  {journal} {\bibinfo  {journal} {MNRAS}\ }\textbf {\bibinfo
  {volume} {505}},\ \bibinfo {pages} {1441} (\bibinfo {year} {2021})},\ \Eprint
  {http://arxiv.org/abs/2105.04025} {arXiv:2105.04025 [astro-ph.CO]}
  \BibitemShut {NoStop}%
\bibitem [{\citenamefont {{Ch{\'a}vez}}\ \emph {et~al.}(2012)\citenamefont
  {{Ch{\'a}vez}}, \citenamefont {{Terlevich}}, \citenamefont {{Terlevich}},
  \citenamefont {{Plionis}}, \citenamefont {{Bresolin}}, \citenamefont
  {{Basilakos}},\ and\ \citenamefont {{Melnick}}}]{Chavez2012}%
  \BibitemOpen
  \bibfield  {author} {\bibinfo {author} {\bibfnamefont {R.}~\bibnamefont
  {{Ch{\'a}vez}}}, \bibinfo {author} {\bibfnamefont {E.}~\bibnamefont
  {{Terlevich}}}, \bibinfo {author} {\bibfnamefont {R.}~\bibnamefont
  {{Terlevich}}}, \bibinfo {author} {\bibfnamefont {M.}~\bibnamefont
  {{Plionis}}}, \bibinfo {author} {\bibfnamefont {F.}~\bibnamefont
  {{Bresolin}}}, \bibinfo {author} {\bibfnamefont {S.}~\bibnamefont
  {{Basilakos}}}, \ and\ \bibinfo {author} {\bibfnamefont {J.}~\bibnamefont
  {{Melnick}}},\ }\href {\doibase 10.1111/j.1745-3933.2012.01299.x} {\bibfield
  {journal} {\bibinfo  {journal} {Monthly Notices of the Royal Astronomical
  Society}\ }\textbf {\bibinfo {volume} {425}},\ \bibinfo {pages} {L56}
  (\bibinfo {year} {2012})},\ \Eprint {http://arxiv.org/abs/1203.6222}
  {arXiv:1203.6222 [astro-ph.CO]} \BibitemShut {NoStop}%
\bibitem [{\citenamefont {{Ch{\'a}vez}}\ \emph {et~al.}(2014)\citenamefont
  {{Ch{\'a}vez}}, \citenamefont {{Terlevich}}, \citenamefont {{Terlevich}},
  \citenamefont {{Bresolin}}, \citenamefont {{Melnick}}, \citenamefont
  {{Plionis}},\ and\ \citenamefont {{Basilakos}}}]{Chavez2014}%
  \BibitemOpen
  \bibfield  {author} {\bibinfo {author} {\bibfnamefont {R.}~\bibnamefont
  {{Ch{\'a}vez}}}, \bibinfo {author} {\bibfnamefont {R.}~\bibnamefont
  {{Terlevich}}}, \bibinfo {author} {\bibfnamefont {E.}~\bibnamefont
  {{Terlevich}}}, \bibinfo {author} {\bibfnamefont {F.}~\bibnamefont
  {{Bresolin}}}, \bibinfo {author} {\bibfnamefont {J.}~\bibnamefont
  {{Melnick}}}, \bibinfo {author} {\bibfnamefont {M.}~\bibnamefont
  {{Plionis}}}, \ and\ \bibinfo {author} {\bibfnamefont {S.}~\bibnamefont
  {{Basilakos}}},\ }\href {\doibase 10.1093/mnras/stu987} {\bibfield  {journal}
  {\bibinfo  {journal} {Monthly Notices of the Royal Astronomical Society}\
  }\textbf {\bibinfo {volume} {442}},\ \bibinfo {pages} {3565} (\bibinfo {year}
  {2014})},\ \Eprint {http://arxiv.org/abs/1405.4010} {arXiv:1405.4010
  [astro-ph.GA]} \BibitemShut {NoStop}%
\bibitem [{\citenamefont {{Terlevich}}\ \emph {et~al.}(2015)\citenamefont
  {{Terlevich}}, \citenamefont {{Terlevich}}, \citenamefont {{Melnick}},
  \citenamefont {{Ch{\'a}vez}}, \citenamefont {{Plionis}}, \citenamefont
  {{Bresolin}},\ and\ \citenamefont {{Basilakos}}}]{Terlevich2015}%
  \BibitemOpen
  \bibfield  {author} {\bibinfo {author} {\bibfnamefont {R.}~\bibnamefont
  {{Terlevich}}}, \bibinfo {author} {\bibfnamefont {E.}~\bibnamefont
  {{Terlevich}}}, \bibinfo {author} {\bibfnamefont {J.}~\bibnamefont
  {{Melnick}}}, \bibinfo {author} {\bibfnamefont {R.}~\bibnamefont
  {{Ch{\'a}vez}}}, \bibinfo {author} {\bibfnamefont {M.}~\bibnamefont
  {{Plionis}}}, \bibinfo {author} {\bibfnamefont {F.}~\bibnamefont
  {{Bresolin}}}, \ and\ \bibinfo {author} {\bibfnamefont {S.}~\bibnamefont
  {{Basilakos}}},\ }\href {\doibase 10.1093/mnras/stv1128} {\bibfield
  {journal} {\bibinfo  {journal} {Monthly Notices of the Royal Astronomical
  Society}\ }\textbf {\bibinfo {volume} {451}},\ \bibinfo {pages} {3001}
  (\bibinfo {year} {2015})},\ \Eprint {http://arxiv.org/abs/1505.04376}
  {arXiv:1505.04376 [astro-ph.CO]} \BibitemShut {NoStop}%
\bibitem [{\citenamefont {{Ch{\'a}vez}}\ \emph {et~al.}(2016)\citenamefont
  {{Ch{\'a}vez}}, \citenamefont {{Plionis}}, \citenamefont {{Basilakos}},
  \citenamefont {{Terlevich}}, \citenamefont {{Terlevich}}, \citenamefont
  {{Melnick}}, \citenamefont {{Bresolin}},\ and\ \citenamefont
  {{Gonz{\'a}lez-Mor{\'a}n}}}]{Chavez2016}%
  \BibitemOpen
  \bibfield  {author} {\bibinfo {author} {\bibfnamefont {R.}~\bibnamefont
  {{Ch{\'a}vez}}}, \bibinfo {author} {\bibfnamefont {M.}~\bibnamefont
  {{Plionis}}}, \bibinfo {author} {\bibfnamefont {S.}~\bibnamefont
  {{Basilakos}}}, \bibinfo {author} {\bibfnamefont {R.}~\bibnamefont
  {{Terlevich}}}, \bibinfo {author} {\bibfnamefont {E.}~\bibnamefont
  {{Terlevich}}}, \bibinfo {author} {\bibfnamefont {J.}~\bibnamefont
  {{Melnick}}}, \bibinfo {author} {\bibfnamefont {F.}~\bibnamefont
  {{Bresolin}}}, \ and\ \bibinfo {author} {\bibfnamefont {A.~L.}\ \bibnamefont
  {{Gonz{\'a}lez-Mor{\'a}n}}},\ }\href {\doibase 10.1093/mnras/stw1813}
  {\bibfield  {journal} {\bibinfo  {journal} {Monthly Notices of the Royal
  Astronomical Society}\ }\textbf {\bibinfo {volume} {462}},\ \bibinfo {pages}
  {2431} (\bibinfo {year} {2016})},\ \Eprint {http://arxiv.org/abs/1607.06458}
  {arXiv:1607.06458 [astro-ph.CO]} \BibitemShut {NoStop}%
\bibitem [{\citenamefont {{Cao, Shuo}}\ \emph {et~al.}(2017)\citenamefont
  {{Cao, Shuo}}, \citenamefont {{Zheng, Xiaogang}}, \citenamefont {{Biesiada,
  Marek}}, \citenamefont {{Qi, Jingzhao}}, \citenamefont {{Chen, Yun}},\ and\
  \citenamefont {{Zhu, Zong-Hong}}}]{ShuoQSO:2017}%
  \BibitemOpen
  \bibfield  {author} {\bibinfo {author} {\bibnamefont {{Cao, Shuo}}}, \bibinfo
  {author} {\bibnamefont {{Zheng, Xiaogang}}}, \bibinfo {author} {\bibnamefont
  {{Biesiada, Marek}}}, \bibinfo {author} {\bibnamefont {{Qi, Jingzhao}}},
  \bibinfo {author} {\bibnamefont {{Chen, Yun}}}, \ and\ \bibinfo {author}
  {\bibnamefont {{Zhu, Zong-Hong}}},\ }\href {\doibase
  10.1051/0004-6361/201730551} {\bibfield  {journal} {\bibinfo  {journal}
  {A\&A}\ }\textbf {\bibinfo {volume} {606}},\ \bibinfo {pages} {A15} (\bibinfo
  {year} {2017})}\BibitemShut {NoStop}%
\bibitem [{\citenamefont {Sandage}(1988)}]{Sandage:1988}%
  \BibitemOpen
  \bibfield  {author} {\bibinfo {author} {\bibfnamefont {A.}~\bibnamefont
  {Sandage}},\ }\href {\doibase 10.1146/annurev.aa.26.090188.003021} {\bibfield
   {journal} {\bibinfo  {journal} {Annual Review of Astronomy and
  Astrophysics}\ }\textbf {\bibinfo {volume} {26}},\ \bibinfo {pages} {561}
  (\bibinfo {year} {1988})},\ \Eprint
  {http://arxiv.org/abs/https://doi.org/10.1146/annurev.aa.26.090188.003021}
  {https://doi.org/10.1146/annurev.aa.26.090188.003021} \BibitemShut {NoStop}%
\bibitem [{\citenamefont {Abbott}\ \emph {et~al.}(2022)\citenamefont {Abbott}
  \emph {et~al.}}]{DES:2021wwk}%
  \BibitemOpen
  \bibfield  {author} {\bibinfo {author} {\bibfnamefont {T.~M.~C.}\
  \bibnamefont {Abbott}} \emph {et~al.} (\bibinfo {collaboration} {DES}),\
  }\href {\doibase 10.1103/PhysRevD.105.023520} {\bibfield  {journal} {\bibinfo
   {journal} {Phys. Rev. D}\ }\textbf {\bibinfo {volume} {105}},\ \bibinfo
  {pages} {023520} (\bibinfo {year} {2022})},\ \Eprint
  {http://arxiv.org/abs/2105.13549} {arXiv:2105.13549 [astro-ph.CO]}
  \BibitemShut {NoStop}%
\bibitem [{\citenamefont {Asgari}\ \emph {et~al.}(2021)\citenamefont {Asgari}
  \emph {et~al.}}]{KiDS:2020suj}%
  \BibitemOpen
  \bibfield  {author} {\bibinfo {author} {\bibfnamefont {M.}~\bibnamefont
  {Asgari}} \emph {et~al.} (\bibinfo {collaboration} {KiDS}),\ }\href {\doibase
  10.1051/0004-6361/202039070} {\bibfield  {journal} {\bibinfo  {journal}
  {Astron. Astrophys.}\ }\textbf {\bibinfo {volume} {645}},\ \bibinfo {pages}
  {A104} (\bibinfo {year} {2021})},\ \Eprint {http://arxiv.org/abs/2007.15633}
  {arXiv:2007.15633 [astro-ph.CO]} \BibitemShut {NoStop}%
\bibitem [{\citenamefont {Heymans}\ \emph {et~al.}(2021)\citenamefont
  {Heymans}, \citenamefont {Tröster}, \citenamefont {Asgari}, \citenamefont
  {Blake}, \citenamefont {Hildebrandt}, \citenamefont {Joachimi}, \citenamefont
  {Kuijken}, \citenamefont {Lin}, \citenamefont {S{\'{a}}nchez}, \citenamefont
  {van~den Busch}, \citenamefont {Wright}, \citenamefont {Amon}, \citenamefont
  {Bilicki}, \citenamefont {de~Jong}, \citenamefont {Crocce}, \citenamefont
  {Dvornik}, \citenamefont {Erben}, \citenamefont {Fortuna}, \citenamefont
  {Getman}, \citenamefont {Giblin}, \citenamefont {Glazebrook}, \citenamefont
  {Hoekstra}, \citenamefont {Joudaki}, \citenamefont {Kannawadi}, \citenamefont
  {Köhlinger}, \citenamefont {Lidman}, \citenamefont {Miller}, \citenamefont
  {Napolitano}, \citenamefont {Parkinson}, \citenamefont {Schneider},
  \citenamefont {Shan}, \citenamefont {Valentijn}, \citenamefont {Kleijn},\
  and\ \citenamefont {Wolf}}]{Heymans_2021}%
  \BibitemOpen
  \bibfield  {author} {\bibinfo {author} {\bibfnamefont {C.}~\bibnamefont
  {Heymans}}, \bibinfo {author} {\bibfnamefont {T.}~\bibnamefont {Tröster}},
  \bibinfo {author} {\bibfnamefont {M.}~\bibnamefont {Asgari}}, \bibinfo
  {author} {\bibfnamefont {C.}~\bibnamefont {Blake}}, \bibinfo {author}
  {\bibfnamefont {H.}~\bibnamefont {Hildebrandt}}, \bibinfo {author}
  {\bibfnamefont {B.}~\bibnamefont {Joachimi}}, \bibinfo {author}
  {\bibfnamefont {K.}~\bibnamefont {Kuijken}}, \bibinfo {author} {\bibfnamefont
  {C.-A.}\ \bibnamefont {Lin}}, \bibinfo {author} {\bibfnamefont {A.~G.}\
  \bibnamefont {S{\'{a}}nchez}}, \bibinfo {author} {\bibfnamefont {J.~L.}\
  \bibnamefont {van~den Busch}}, \bibinfo {author} {\bibfnamefont {A.~H.}\
  \bibnamefont {Wright}}, \bibinfo {author} {\bibfnamefont {A.}~\bibnamefont
  {Amon}}, \bibinfo {author} {\bibfnamefont {M.}~\bibnamefont {Bilicki}},
  \bibinfo {author} {\bibfnamefont {J.}~\bibnamefont {de~Jong}}, \bibinfo
  {author} {\bibfnamefont {M.}~\bibnamefont {Crocce}}, \bibinfo {author}
  {\bibfnamefont {A.}~\bibnamefont {Dvornik}}, \bibinfo {author} {\bibfnamefont
  {T.}~\bibnamefont {Erben}}, \bibinfo {author} {\bibfnamefont {M.~C.}\
  \bibnamefont {Fortuna}}, \bibinfo {author} {\bibfnamefont {F.}~\bibnamefont
  {Getman}}, \bibinfo {author} {\bibfnamefont {B.}~\bibnamefont {Giblin}},
  \bibinfo {author} {\bibfnamefont {K.}~\bibnamefont {Glazebrook}}, \bibinfo
  {author} {\bibfnamefont {H.}~\bibnamefont {Hoekstra}}, \bibinfo {author}
  {\bibfnamefont {S.}~\bibnamefont {Joudaki}}, \bibinfo {author} {\bibfnamefont
  {A.}~\bibnamefont {Kannawadi}}, \bibinfo {author} {\bibfnamefont
  {F.}~\bibnamefont {Köhlinger}}, \bibinfo {author} {\bibfnamefont
  {C.}~\bibnamefont {Lidman}}, \bibinfo {author} {\bibfnamefont
  {L.}~\bibnamefont {Miller}}, \bibinfo {author} {\bibfnamefont {N.~R.}\
  \bibnamefont {Napolitano}}, \bibinfo {author} {\bibfnamefont
  {D.}~\bibnamefont {Parkinson}}, \bibinfo {author} {\bibfnamefont
  {P.}~\bibnamefont {Schneider}}, \bibinfo {author} {\bibfnamefont
  {H.}~\bibnamefont {Shan}}, \bibinfo {author} {\bibfnamefont {E.~A.}\
  \bibnamefont {Valentijn}}, \bibinfo {author} {\bibfnamefont {G.~V.}\
  \bibnamefont {Kleijn}}, \ and\ \bibinfo {author} {\bibfnamefont
  {C.}~\bibnamefont {Wolf}},\ }\href {\doibase 10.1051/0004-6361/202039063}
  {\bibfield  {journal} {\bibinfo  {journal} {Astronomy {\&} Astrophysics}\
  }\textbf {\bibinfo {volume} {646}},\ \bibinfo {pages} {A140} (\bibinfo {year}
  {2021})}\BibitemShut {NoStop}%
\end{thebibliography}%

\end{document}